\documentclass[10pt,journal]{IEEEtran}
\ifCLASSINFOpdf
\else
\fi
\usepackage{amsmath}
\usepackage{graphicx}
\usepackage{tabularx}
\usepackage{color}
\usepackage{balance,subfigure}
\usepackage{multirow}
\usepackage{rotating}
\usepackage{epstopdf}
\usepackage{wrapfig}
\usepackage{array}
\usepackage{cite}

\def \etal {et al.}
\begin{document}
%
\title{Semi-Supervised Speech Emotion Recognition with Ladder Networks}
%
%
%

\author{Srinivas~Parthasarathy,~\IEEEmembership{Student Member,~IEEE,}
	Carlos~Busso,~\IEEEmembership{Senior Member, IEEE,}
	
	\thanks{S. Parthasarathy was with the Erik Jonsson School of Engineering \& Computer Science, The University of Texas at Dallas, TX 75080 USA (e-mail: sxp120931@utdallas.edu).}
	\thanks{C. Busso is with the Erik Jonsson School of Engineering \& Computer Science, The University of Texas at Dallas, TX 75080 USA (e-mail: busso@utdallas.edu).}
	\thanks{Manuscript received May 7, 2019; revised xxxx xx, xxxx.}}   

%
%

\markboth{IEEE/ACM Transactions on Audio, Speech, and Language Processing,~Vol.~xx, No.~x, May~2019}%
{Shell \MakeLowercase{\textit{et al.}}: Bare Demo of IEEEtran.cls for IEEE Journals}
%



\maketitle

\begin{abstract}
\emph{Speech emotion recognition} (SER) systems find applications in various fields such as healthcare, education, and security and defense. A major drawback of these systems is their lack of generalization across different conditions. For example, systems that show superior performance on certain databases show poor performance when tested on other corpora. This problem can be solved by training models on large amounts of labeled data from the target domain, which is expensive and time-consuming. Another approach is to increase the generalization of the models. An effective way to achieve this goal is by regularizing the models through \emph{multitask learning} (MTL), where auxiliary tasks are learned along with the primary task. These methods often require the use of labeled data which is computationally expensive to collect for emotion recognition (gender, speaker identity, age or other emotional descriptors). This study proposes the use of ladder networks for emotion recognition, which utilizes an unsupervised auxiliary task. The primary task is a regression problem to predict emotional attributes. The auxiliary task is the reconstruction of intermediate feature representations using a denoising autoencoder. This auxiliary task does not require labels so it is possible to train the framework in a semi-supervised fashion with abundant unlabeled data from the target domain. This study shows that the proposed approach creates a powerful framework for SER, achieving superior performance than fully supervised \emph{single-task learning} (STL) and MTL baselines. The approach is implemented with several acoustic features, showing that ladder networks generalize significantly better in cross-corpus settings. Compared to the STL baselines, the proposed approach achieves relative gains in \emph{concordance correlation coefficient} (CCC) between 3.0\% and 3.5\% for within corpus evaluations, and between 16.1\%  and 74.1\% for cross corpus evaluations, highlighting the power of the architecture.

\end{abstract}


%
\IEEEpeerreviewmaketitle

\section{Introduction}
\label{sec:intro}

Recognizing emotions is a key feature needed to build socially aware systems. Therefore, it is an important part of \emph{human computer interaction} (HCI). Emotion recognition can play an important role in various fields such as healthcare (mood profiles) \cite{Cummins_2015}, education (tutoring) \cite{Litman_2004} and security and defense (surveillance)  \cite{Clavel_2008}. \emph{Speech emotion recognition} (SER) have enormous potential given the ubiquity of speech-based devices. However, it is important that SER models generalize well across different conditions and settings showing robust performance.

Conventionally, emotion recognition systems are trained with supervised learning solutions. The generalization of the models is often emphasized by training on a variety of samples with diverse labels \cite{Zhang_2017}. The state-of-the-art models for standard computer vision tasks utilize thousands of labeled samples. Similarly, \emph{automatic speech recognition} (ASR) systems are trained on several hundred hours of data with transcriptions. Generally, labels for emotion recognition tasks are collected with perceptual evaluations from multiple evaluators. The raters annotate samples by listening or watching to the stimulus. This evaluation procedure is cognitively intense and expensive. Therefore, standard benchmark datasets for SER have limited number of sentences with emotional labels, often collected from a limited number of evaluators. This limitation severely affects the generalization of the systems.

An alternative approach to increase the generalization of the models is by building robust models. An effective approach to achieve this goal is with \emph{multitask learning} (MTL) \cite{Caruana_1997}, where relevant auxiliary tasks are simultaneously solved along with the primary task. By solving relevant auxiliary tasks, the models are regularized by finding more general high-level feature representations that are still discriminative for the primary task. Multitask learning has been successfully used for emotion recognition tasks \cite{Lotfian_2018, Parthasarathy_2017_3, Xia_2017,Tao_2018}. While these MTL methods have achieved promising results, most of the proposed solutions have focused on MTL problems that utilize supervised auxiliary tasks. Examples include gender recognition \cite{Kim_2017_2,Tao_2018}, speaker information \cite{Tao_2018}, other emotional attributes \cite{Parthasarathy_2017_3, Xia_2017} and secondary emotions \cite{Lotfian_2018}. This approach requires the use of meta labels which further limits the training of the models. In many scenarios, it is possible to collect large amount of data without labels from the target domain. It is important to build models that can effectively utilize unsupervised auxiliary tasks to regularize the model, leveraging these unlabelled recordings. This study explores this idea with ladder networks, building upon our previous work \cite{Parthasarathy_2018_3}. The ladder network architecture is a framework that combines supervised tasks with unsupervised auxiliary tasks. These auxiliary tasks correspond to the reconstruction of feature representations at various layers in a \emph{deep neural network} (DNN). In essence, the ladder network is a denoising autoencoder trained along with a supervised classifier or regressor that utilizes the encoded representation. The uniqueness of the framework is the skip connections between the corresponding encoder and decoder layers, which reduce the load on the encoder layers to carry information for decoding the layers. With this approach, the higher layers of the encoder learn discriminative representations for the supervised task. Furthermore, the reconstruction of the feature representations is completely unsupervised. This aspect of the model enables us to use the framework in a semi-supervised fashion, leveraging large amount of data from the target domain without emotional labels.

This study provides a comprehensive analysis of auxiliary tasks for speech emotion recognition on the MSP-Podcast corpus \cite{Lotfian_201x}. The study focuses on regression problems, where the primary task is to predict the arousal, valence and dominance scores. The proposed implementation uses \emph{high-level descriptors} (HLDs), computed at the speaking turn-level as feature inputs. The evaluation compares the performance of the proposed ladder network framework for emotion recognition against \emph{single-task learning} (STL) and MTL baselines. The proposed framework is tested in a fully supervised setting as well as in  a semi-supervised setting using unlabeled data (around 10 times the amount of labeled data in the corpus). The experimental results show the benefits of the framework, obtaining state-of-the-art performance on this speech emotional corpus. This study also examines the performance of the proposed architectures for cross-corpus experiments, where the models are trained on the MSP-Podcast corpus and tested on other popular databases for SER tasks (USC-IEMOCAP and MSP-IMPROV corpora). The proposed architectures achieve significant improvements in the cross-corpus experiments, validating the regularizing power of the proposed framework, leading to models that generalize better to unseen conditions. The gains in performance are particularly clear when the approach is implemented in a semi-supervised fashion, adapting the models using unlabeled data from the target domain. Finally, this study replicates the proposed architecture for two frame-level feature inputs: (a) dynamic \emph{low-level descriptors} (LLDs), and (b) \emph{Mel-frequency band} (MFB) energies. While the HLDs are sentence-level feature representation, the LLDs and MFBs are frame-level feature representations. The model with ladder networks achieve significant improvements over the baselines in most cases. 

The rest of the paper is organized as follows. Section \ref{sec:background} reviews studies on research areas that are relevant to this work. Section \ref{sec:methodology} presents the proposed architecture that exploits unsupervised auxiliary tasks to regularize the network. Section \ref{sec:exp_setup} gives details on the experimental setup including the databases and features used in this study. Section \ref{sec:results} presents the exhaustive experimental evaluations, showing the benefits of the proposed architecture. Finally, Section \ref{sec:conclusion} provides the concluding remarks, discussing potential areas of improvements.

\section{Background}
\label{sec:background}

This study uses the emotional attributes arousal, valence and dominance to describe emotions. SER systems for these problems are often built to recognize individual emotional attributes. Most frameworks are trained in a supervised fashion with labeled data. Given the limited size of most speech emotional databases, these supervised emotion recognition frameworks are commonly trained with a few hours of labeled data. Using unlabeled data is an interesting method to increase the generalization of the models to a new domain. 

\subsection {Semi-Supervised Learning}
\label{ssec:semi}

Previous studies for semi-supervised learning have considered the \emph{inductive learning} technique,  where a classifier is first trained on the labeled samples. The trained classifier is then used on the unlabeled set to obtain predictions. The training set is then augmented with samples having highly confident predictions. The classifier is retrained with this augmented training set. This process is iterated a fixed number of times after which the performance often saturates. Zhang \etal \cite{Zhang_2011_2} used this inductive learning procedure for SER to leverage unlabeled data. They enhanced their supervised learning approach with this method, obtaining better predictions on labeled data \cite{Zhang_2016, Zhang_2018_3}. Cohen \etal \cite{Cohen_2003_3, cohen_2003_4} proposed a similar strategy for facial expressions using probabilistic Bayesian classifiers. 

Another approach for semi-supervised learning is the \emph{co-training} or \emph{multi-view learning} procedure \cite{Blum_1998}. In this method, the classifiers are trained on distinct feature partitions (views). The different classifiers are used for predictions on the unlabeled data, augmenting the training set with samples that are consistently recognized by the classifiers. Mahdhaoui and Chetouani \cite{Mahdhaoui_2010} proposed multi-view training for SER using different sets of acoustic features. Similarly, Zhang \etal \cite{Zhang_2015} utilized co-training along with \emph{active learning} where they only annotated emotional labels for samples where the predictions were made with low confidence by the multi-view classifiers. Liu \etal \cite{Liu_2007} proposed multi-view learning for SER, where they used temporal and statistical acoustic features. Studies have also considered multi-view learning by incorporating multiple modalities \cite{Zhang_2018_3, Zhang_2016, Kim_2017_2}.

This study is more closely related to the recent advances in deep learning that combine supervised and unsupervised learning. Similar to our work, Deng \etal \cite{Deng_2018} proposed combining an autoencoder and a classifier for SER. Their framework is based on a discriminative \emph{Restricted Boltzmann Machine} (RBM), which considers unlabeled samples as an extra \emph{garbage} class in the classification problem. Huang \etal \cite{Huang_2014} proposed learning affect sensitive features using a semi-supervised implementation of a \emph{convolutional neural network} (CNN) for SER. In this study, general features are learned using an unsupervised CNN architecture, and then these features are fine-tuned for affect recognition. Similarly, Mao \etal \cite{Mao_2014} trained a CNN to learn salient features for SER. The CNNs were first trained on unlabeled samples using a sparse autoencoder and a reconstruction penalization. The invariant features were then used as inputs for learning affect sensitive feature representations. Our work follows these studies, further extending semi-supervised SER. Our study differs from previous studies by effectively training an autoencoder and a regressor together, such that the auxiliary task of reconstructing the input feature vector and intermediate feature representations helps the primary supervised learning task. Jointly training the autoencoder (auxiliary task) and the regression problem (primary task) is an important contribution leading to more discriminative SER models.

\subsection{Auxiliary Tasks and Multitask Learning}
\label{ssec:AuxiliaryReview}

There are multiple studies that have analyzed the regularizing benefits of auxiliary tasks for SER. Xia and Liu \cite{Xia_2017} combined the learning of emotional categories and emotional attributes. The primary task was the classification of emotional categories. The secondary task was either classification or regression of emotional attributes. Parthasarathy and Busso \cite{Parthasarathy_2017_3} proposed to jointly predict arousal, valence and dominance scores using a MTL framework, where recognizing one of the attributes was the primary task and recognizing the other two attributes were the secondary tasks. The MTL framework learned the inherent correlation between the various emotional attributes leading to improvements over STL. Similarly, Chang and Scherer \cite{Chang_2017} used arousal prediction as an auxiliary task for a valence classifier. Chen \etal \cite{Chen_2017} showed similar improvements in performance for the prediction of time-continuous emotional attributes. Their system jointly predicted arousal and valence scores, obtaining the best performance for the affect sub-task in the \emph{audio/visual emotion challenge} (AVEC) in 2017 \cite{Ringeval_2017}. Le \etal \cite{Le_2017} also used a MTL framework for time-continuous attribute recognition. Their framework trained classifiers by discretizing attribute scores into discrete classes using the \emph{k-means} algorithm with $k\in \{4, 6, 8, 10\}$. The different classifiers were then learned together as multiple auxiliary tasks using MTL framework. (e.g., learning together a four-class problem and a six-class problem). Similarly, Lotfian and Busso \cite{Lotfian_2018} showed improvements for categorical emotion recognition by using a MTL framework for learning the dominant emotion (primary task) and secondary emotions also conveyed in the sentence (auxiliary task). 

Previous studies have also considered using other auxiliary tasks to improve SER. Kim \etal \cite{Kim_2017} used gender and naturalness recognition as auxiliary tasks for emotion recognition. The naturalness task consisted of a binary classifier that determines whether the sentences were natural or acted recordings across different databases. Tao and Liu \cite{Tao_2018} used gender recognition and speaker identification as auxiliary tasks for classifying emotional categories on the USC-IEMOCAP corpus. Similarly, Zhao \etal \cite{Zhang_2018_3} transferred age and gender attributes as auxiliary tasks to predict emotion attributes. Abdelwahab and Busso \cite{Abdelwahab_2018_3} used an auxiliary task for cross-corpus SER. The auxiliary task learned common representation between the source and target domains using a \emph{domain adversarial neural network} (DANN).

\subsection{Ladder Networks}
\label{ssec:reviewladder}

The idea of ladder networks was first proposed by Valpola \cite{Valpola_2015}. This work showed the benefits of using lateral shortcut connections to aid deep unsupervised learning. Rasmus \etal \cite{Rasmus_2015,Rasmus_2015_2} further extended this idea to support supervised learning. Classification and regression tasks were added to the unsupervised reconstruction of inputs through a denoising autoencoder. Finally, Pezeshki \etal \cite{Pezeshki_2016} studied the various components that affected the ladder network, noting that lateral connections between encoder and decoder and the addition of noise at every layer of the network greatly contributed to the improved performance of this framework.

The use of ladder network for SER is appealing since the auxiliary task is unsupervised, so we can use data from the target domain without labels. This feature of the proposed approach is a key distinction between our work and most MTL studies, which use supervised auxiliary tasks. The closest study to our paper is the work of Huang \etal \cite{Huang_2018_2}, which was simultaneously developed with our preliminary study \cite{Parthasarathy_2018_3}. They also proposed ladder networks for SER tasks. A key distinction between this study and our paper is that Huang \etal \cite{Huang_2018_2} only used ladder network to learn feature representations, where the final classifier was a separate \emph{support vector machine} (SVM). This two-step process is equivalent to training an autoencoder followed by a classifier. Instead, our proposed formulation jointly optimizes the unsupervised reconstruction and the supervised regression task in a single step. The proposed ladder network provides the final predictions for the emotional attribute without any additional regressor, which (1) creates better feature representations that are discriminative for the target task, (2) allows our formulation to be trained as an end-to-end system (Sec. \ref{ssec:cnn}).

\section{Proposed Methodology}
\label{sec:methodology}
\subsection{Motivation}
\label{ssec:motivation}

As stated in Section \ref{sec:background}, data with emotional labels are limited. Furthermore, data from the source domain (train set) is not guaranteed to have the same distribution as the target domain (test set). Therefore, most supervised frameworks trained on one corpus do not generalize well when tested across different tasks and corpora. Therefore, there is a fundamental need to regularize architectures such that they generalize across different tasks. This study aims to increase the generalization of SER models with (1) unsupervised auxililary tasks, (2) and unlabeled data from the target domain. Our motivations are based on solving unsupervised auxiliary tasks, which aid the primary emotion recognition task. First, we want to fully utilize available labeled data which is expensive to annotate. To this extent, we build a MTL framework (Section \ref{ssec:mtl_method}) where we jointly learn the dependencies between multiple emotional attributes. The MTL framework regularizes our architecture, but it still demands the utilization of expensive data labeled with emotional information. While labeling audio data for emotion is expensive, unlabeled data is more easily available. The amount of unlabeled data is often greater than the amount of labeled data. The unlabeled data from the target domain can be used to reduce the gap between the source and target domains. We propose to use ladder networks (Section \ref{ssec:ladder_method}) to effectively leverage unlabeled data. Collectively, the MTL approach combined with the ladder network creates a semi-supervised architecture that effectively generalizes to new domains. This study shows that these powerful representations created by our model can be used across emotional corpora to achieve state-of-the-art SER performance.

\begin{figure}[tb]
	\centering
	\includegraphics[width=\columnwidth]{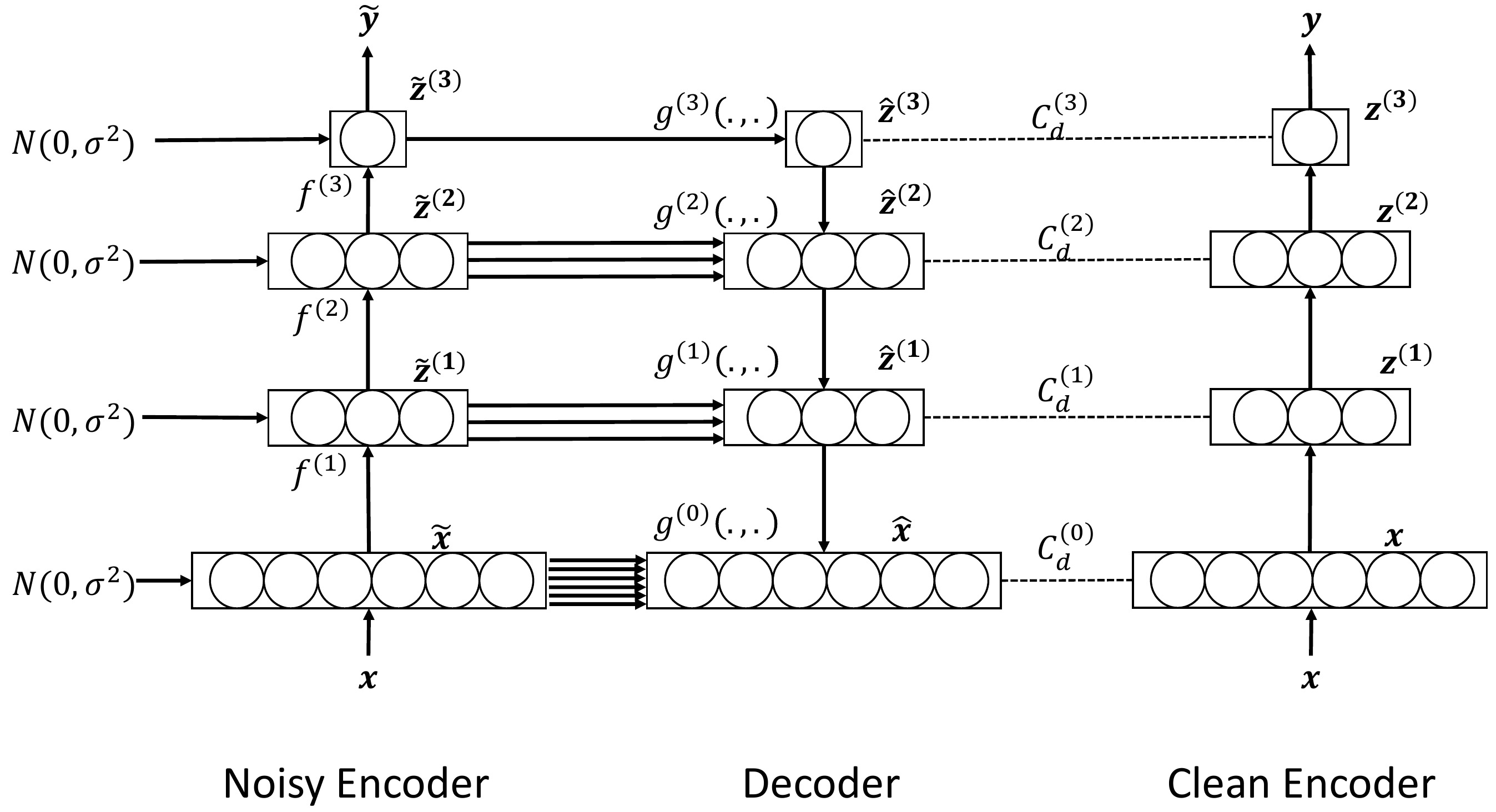}
	\caption{Ladder network architecture using auxiliary tasks for emotion attribute prediction. The network has a noisy encoder, decoder and clear encoder used for inferences. The ladder connections connect the noisy encoder with the decoder.}
	\label{fig:ladder}
\end{figure}

\subsection{Ladder Network for Speech Emotion Recognition}
\label{ssec:ladder_method}
Ladder networks, at their core, combine an unsupervised auxiliary task with a supervised classifier or regressor. Using an autoencoder for supervised tasks is not new. Traditionally, the autoencoder is trained separately from the supervised task, where its goal is to learn features representations that are useful for reconstructing the input. However, the information needed to reconstruct the input does not necessarily create a discriminative representation for the classification or regression task. Therefore, it is important to combine the training of the autoencoder with the supervised task, which is a key feature of the ladder networks. Figure \ref{fig:ladder} illustrates a conceptual ladder network. A noisy version of the encoder is created by adding noise at every layer of the encoder. The goal of the autoencoder is to reconstruct the feature representations at the input and intermediate layers. The core concept of the autoencoder in the ladder network involves skip connections between corresponding encoder and decoder layers. Effectively, these skip connections provide a shortcut between the decoder and encoder, bypassing higher layers of the encoder. Therefore, the top layers of the encoder can learn representations better suited towards the primary discriminative task. This is a fundamental difference with simple autoencoders. Note that the ladder network combines the supervised task with an unsupervised auxiliary task. Therefore, the true benefit of the architecture is when it is used in a semi-supervised fashion. The rest of this section explains in detail the encoder and decoder of the ladder network.  

\noindent\textbf{Encoder:} 
The encoder consists of a \emph{multilayer perceptron} (MLP). A zero-mean Gaussian noise with variance $\sigma^2$ is added to each layer of the MLP ($N(0, \sigma^2)$ in Fig. \ref{fig:ladder}). The decoder is constructed to denoise the noisy latent representations $\tilde{\mathbf{z}}$ at every layer. Therefore, a clean copy of the encoder path is built to get the targets $\mathbf{z}$ for reconstruction (\emph{clean encoder} in Fig. \ref{fig:ladder}). Since the architecture reconstructs intermediate layers, $\hat{\mathbf{z}}$, a trivial solution to minimize the cost is $\hat{\mathbf{z}}$ = $\mathbf{z}$ = constant. To avoid this trivial solution, intermediate layers are normalized using batch normalization. Batch normalization is performed on all layers except the input layer. The scaling and bias values are learned as trainable parameters before applying the activation. Besides encoding the representations for reconstruction, the final layer of the encoder, $\mathbf{\tilde{z}^{(L)}}$, is used for training the supervised regression task, which in our case is the prediction of emotional attributes. The noisy representation $\tilde{\mathbf{z}}$ further regularizes the network. The clean representations $\mathbf{z}$ are used during inference.

\noindent\textbf{Decoder:} 
Similar to the encoder, the decoder of the ladder network is a MLP (\emph{decoder} in Fig. \ref{fig:ladder}). The layers of the decoder network mirrors the layers of the encoder. The decoder is constructed to denoise the noisy representations of the encoder. The denoising process combines top down information from the decoder ($\mathbf{\hat{z}^{(l+1)}}$) with lateral information from the corresponding encoder layer ($\mathbf{\tilde{z}^{(l)}}$). With the lateral connections, the network passes the information needed for denoising the latent representations, bypassing the top layers of the encoder, which can, instead, provide abstract discriminative information for the supervised regression task. As a result, an unsupervised auxiliary cost is added without sacrificing the performance of the architecture for the supervised task. Different denoising functions, $\mathbf{g(\cdot)}$, can model different probability distributions of the latent variables \cite{Rasmus_2015, Pezeshki_2016}. Previous studies have shown that a single layer MLP combining top decoder layers and lateral encoder layers works the best for most tasks. Our preliminary experiments concluded that the same observation also holds for SER tasks. The denoising function $\mathbf{g(\cdot)}$ takes as input $\mathbf{u, \tilde{z}}$ and $\mathbf{u \odot z}$ (layer abbreviations are dropped for clarity), where $\mathbf{u}$ is a batch normalized projection of the decoder layer above $\mathbf{\hat{z}^{(l+1)}}$,  $\mathbf{\tilde{z}}$ is the corresponding noisy representation, and $\mathbf{u \odot z}$ is a element wise product between the decoder and encoder elements. The element wise multiplication assumes that the latent variables are conditionally independent and modulates the encoder representation with the previous decoder layer ($\mathbf{\hat{z}^{(l+1)}}$).

The overall loss for the ladder network is given by 
\begin{align}
C_{Ladder} = C_c + \lambda_l \sum_l C_d^{(l)}
\label{eq:CostLadder}
\end{align}

\noindent where $C_d^{(l)}$ is the reconstruction loss at layer $l$ and $\lambda_l$ is a hyper-parameter that weighs the reconstruction loss at that layer. The supervised loss for predicting the emotional attributes, $C_c$, is added when labeled samples are available. Section \ref{ssec:architecture} gives the implementation and experimental setup used to train the regressor using the proposed ladder network framework.

\subsection{Multitask Learning for Emotion Recognition}
\label{ssec:mtl_method}

\begin{figure}[tb]
	\centering
	\includegraphics[width=2.5cm]{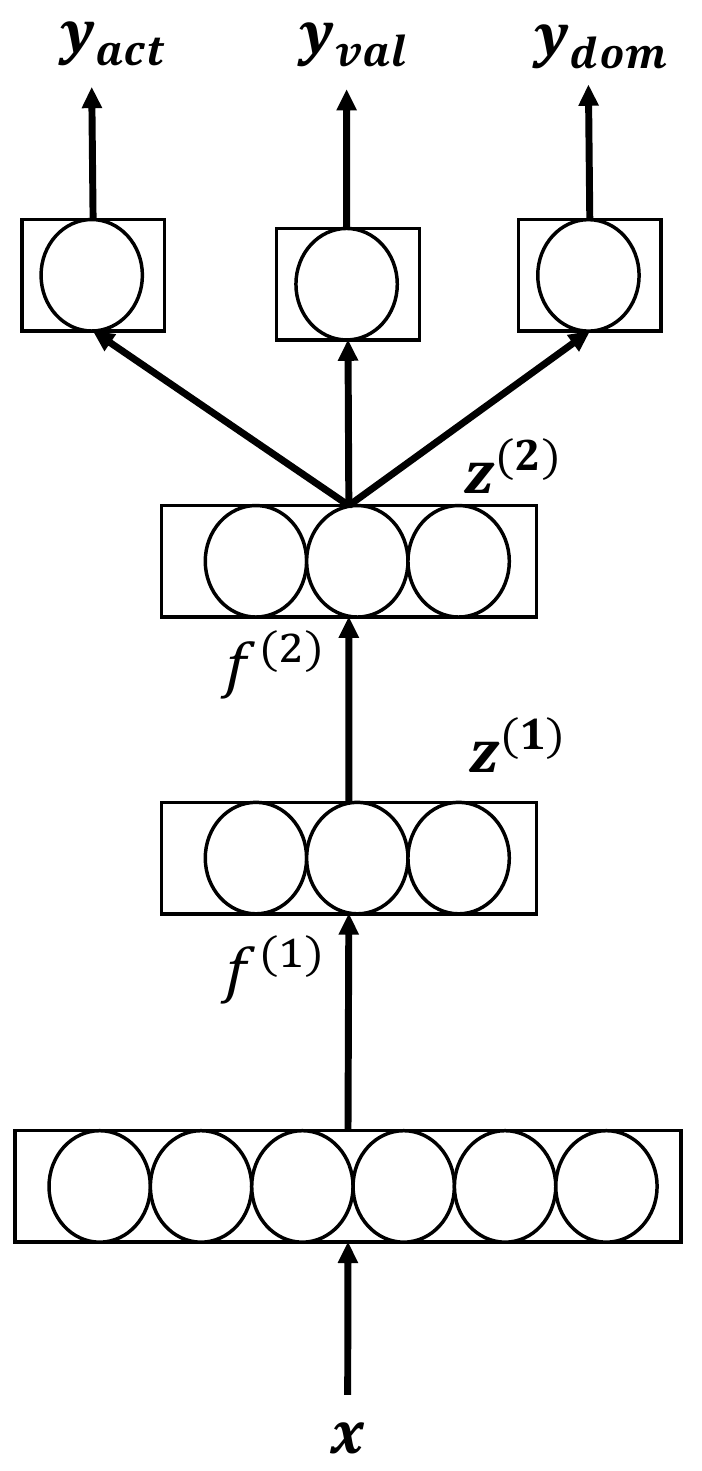}

	\caption{\emph{Multitask learning} (MTL)  architecture to jointly predict arousal, valence and dominance \cite{Parthasarathy_2017_3}. The ladder network architecture can be implemented with MTL, combining supervised and unsupervised auxiliary tasks.}
    	\label{fig:mtl}
\end{figure}

While the ladder network architecture makes efficient use of unlabeled samples to regularize the models, the generalization of the models can also be achieved by better utilizing labeled samples. For the prediction of emotional attributes, one appealing method is to jointly learn multiple emotional attributes. This procedure can be effectively done through MTL with shared and attribute-dependent layers \cite{Parthasarathy_2017_3}. Figure \ref{fig:mtl} illustrates a MTL network with shared hidden layers that jointly predicts arousal, valence and dominance scores. The overall loss for the MTL architecture is given by

\begin{align}
C_{\mathit{MTL}} = \alpha C_{\mathit{aro}} + \beta C_{\mathit{val}} + (1 - \alpha - \beta) C_{\mathit{dom}},
\label{eq:costMTL}
\end{align}

\noindent where $C_{\mathit{aro}}$, $C_{\mathit{val}}$ and $C_{\mathit{dom}}$ are individual losses for the prediction of arousal, valence and dominance, respectively. These losses are multiplied by the hyper-parameters $\alpha$ and $\beta$, respectively, with $\alpha, \beta \in [0,1]$ and $\alpha + \beta \leq 1$. Particular solutions of this formulation are the STL frameworks for arousal ($\alpha=1$, $\beta=0$), valence ($\alpha=0$, $\beta=1$) and dominance ($\alpha=0$, $\beta=0$).

An interesting extension of the proposed ladder network formulation for SER is combining the unsupervised and supervised auxiliary losses. We achieve this goal by replacing $C_c$ in Equation \ref{eq:CostLadder} with $C_{\mathit{MTL}}$ from Equation \ref{eq:costMTL}. In Section \ref{sec:results}, we evaluate the implementation of the ladder network with STL and MTL.

\begin{align}
C_{Lad+MTL} = \alpha C_{\mathit{aro}} + \beta C_{\mathit{val}} & + (1 - \alpha - \beta) C_{\mathit{dom}} \nonumber \\
& + \lambda_l \sum_l C_d^{(l)} \label{eq:CostLadder_MTL}
\end{align}

\section{Experimental Setup}
\label{sec:exp_setup}

\subsection{Datasets}
\label{ssec:data}

This study uses multiple datasets for the different experiments in Section \ref{sec:results}. The primary corpus is the MSP-Podcast (Version 1.2) \cite{Lotfian_2017}, used for all the within corpus experiments (Sec. \ref{ssec:within_corpus}). The MSP-Podcast contains speech collected from online downloadable audio shows, covering various topics such as politics, sports, entertainment, and motivation talks. Therefore, they contain naturalistic speech spanning the emotional spectrum observed during natural conversations. We use a diarization toolkit which identifies segments from distinct speakers. The podcast conversations are sequentially analyzed by automatic algorithms to remove music, silence portions and noisy recordings. We also remove segments with overlapped speech. The selected segments contain a single speaker with duration between 2.75s and 11s. To balance the emotional content of the corpus, we retrieve samples that we believe are emotional following the idea proposed in Mariooryad \etal \cite{Mariooryad_2014_3}. Overall, the corpus contains 50 hours of speech (29,440 speaking turns), which were annotated with emotional labels using Amazon Mechanical Turk. The perceptual evaluation used a modified version of the crowdsource-based protocol presented in Burmania \etal \cite{Burmania_2016_2} to track in real-time the performance of the annotators. The data was annotated for both categorical emotions as well as emotional attributes. This study focuses on the emotional attributes. Each speaking turn was annotated on a scale from one to seven by at least five raters for arousal (1 - very calm, 7 - very active), valence (1 - very negative, 7 - very positive) and dominance (1 - very weak, 7 - very strong). We manually identified speaking turns belonging to 346 speakers in the MSP-Podcast database. The test set contains data from 50 speakers (7,341 speaking turns). The development set contains data from 20 speakers (3,753 speaking turns). The training set has the remaining labeled speaking turns (18,346 segments). This data partition aims to create speaker independent sets for the train, development and testing sets. Besides the labeled data, the MSP-Podcast also contains more than 300 hours of unlabeled data (175,196 segments), corresponding to the pool of clean segments identified from the podcasts, which have not been annotated. This study uses these segments to train the ladder networks (Sec. \ref{ssec:ladder_method}) in a semi-supervised fashion. Section \ref{ssec:within_corpus} presents the results of the experiments conducted on the MSP-Podcast corpus.

Besides the MSP-Podcast corpus, we use two other databases for cross corpora evaluations (Sec. \ref{ssec:cross_corpora}). The first database is the USC-IEMOCAP corpus \cite{Busso_2008}, which contains interactions between pairs of actors improvising scenarios. The database contains 10,527 speaking turns from 10 actors appearing in five dyadic sessions. The speech segments were annotated for arousal, valence and dominance by two raters on a five-Likert scale. More information about this corpus is provided in Busso \etal \cite{Busso_2008_5}. We also use the MSP-IMPROV corpus \cite{Busso_2017}, which contains interactions between pairs of actors improvising scenarios. In addition to the improvised scenarios, the dataset also contains the interactions between the actors during the breaks, resulting in more naturalistic data. The MSP-IMPROV corpus was annotated with emotional labels using Amazon Mechanical Turk using the approach proposed by Burmania \etal \cite{Burmania_2016_2}. Each sentence was annotated for arousal, valence and dominance by five or more raters using a five-Likert scale. More information about this corpus is provided in Busso \etal \cite{Busso_2017}.

\subsection{Acoustic Features}
\label{ssec:features}
This study predominantly uses the acoustic features introduced for the paralinguistic challenge at Interspeech 2013 \cite{Schuller_2013}. These features, which are referred to as the ComParE feature set, are extracted in a two-step procedure. First, LLDs are extracted over 20 millisecond frames (100 fps). These LLDs include loudness, \emph{mel-frequency cepstral coefficients} (MFCCs), fundamental frequency (F0), spectral flux, spectral slope, jitter and shimmer. Second, segment-level features are calculated over the LLDs, leading to a fixed dimensional feature vector. These statistics are referred to as \emph{high-level descriptors} (HLDs) and include various functionals such as the arithmetic and geometric means, standard deviations, peak to peak distances and rise and fall times. The ComParE feature set contains 130 LLDs (65 LLDs + 65 delta) and 6,373 HLDs. Most databases are annotated at the segment-level with a single annotation capturing the emotional content of the entire segment. Since speech segments have variable lengths, most emotion recognition algorithms have to deal with variable length inputs. The HLDs alleviate this problem by creating a fixed dimension input regardless of the length of the sequence. Previous studies have shown the benefits of HLDs for SER tasks \cite{Wollmer_2010_2, Parthasarathy_2017_3}.

\subsection{System Description}
\label{ssec:description}

This study uses two baselines and different implementations of the proposed approach to analyze the performance of the ladder network architecture. All regression systems are trained on the train set, optimizing their performances on the development set. The best system per condition in the development set is then evaluated on the test set, where we report the results.

The study uses two baselines to compare the performance of the proposed architecture. The first baseline uses the STL framework, which is the conventional method for the regression of emotional attributes. The STL framework considers only one of the emotional attribute at a time, creating separate models for arousal, valence and dominance. This approach is referred to as \emph{STL}. The second baseline uses the MTL framework proposed by Parthasarathy and Busso \cite{Parthasarathy_2017_3} (Section \ref{ssec:mtl_method}). This system jointly predicts all three emotional attributes, but it only uses supervised auxiliary tasks without the ladder network. It is expected that the MTL systems should provide a stronger baseline compared to the STL systems, since they use supervised auxiliary tasks. This approach is referred to as \emph{MTL}.

The ladder network architecture, denoted with \emph{Lad}, is studied using four implementations grouped into two settings. The first setting only uses the labeled portion of the corpus. The ladder network is implemented as a supervised problem. We denote this setting by adding the term \emph{L} to the name of the system. The second setting uses the entire corpus containing the labeled and unlabeled portions of the corpus. The ladder network is implemented as a semi-supervised problem. For training with the unlabeled set, we alternate between a mini-batch of unlabeled samples and a mini-batch of labeled samples. We denote this setting by adding the term \emph{UL} to the name of the system.  For both settings, we implement the ladder network with either STL (Eq. \ref{eq:CostLadder}) or MTL (Eq. \ref{eq:CostLadder_MTL}). We denote the corresponding implementation by adding the term  \emph{STL} or \emph{MTL} to the name of the system. For example, the ladder network trained with labeled and unlabeled data using STL is denoted as \emph{Lad + UL + STL}. We expect that combining MTL with the ladder network should result in improved performance as we use both supervised and unsupervised auxiliary tasks to aid our primary task of predicting emotional attributes.

\subsection{System Architecture}
\label{ssec:architecture}

The baselines and the proposed ladder network models are implemented with feed forward dense networks using sentence-level features as inputs. The dense networks contain two hidden layers with 256 nodes in each layer. The activation of the neurons in each layer corresponds to the \emph{rectified linear unit} (ReLU). The input to the dense network is a 6,373D feature vector containing the HLDs for a speaking turn (Sec. \ref{ssec:features}). The output is the predicted value of the emotional attribute. The features and labels are normalized using the z-normalization with the mean and standard deviation calculated over the train set. The models are trained with a learning rate of $5e-5$ for 100 epochs. The model with the best performance on the development set across epochs is evaluated on the test set.

For the architectures of the STL and MTL baselines, we include a dropout of $p=0.5$ between  the  input and the first hidden layer, and between the first and second hidden layers. This setting provides the best regularization on the development set. The hyper-parameters for the MTL methods are optimized using the development set. The parameters $\alpha$ and $\beta$ are separately optimized for each emotional attribute using the development set. Therefore, we have three systems, one for each attribute, with different combination for $\alpha$ and $\beta$.

For the ladder network, we only use dropout between the input layer and the first hidden layer,  following our previous work \cite{Parthasarathy_2018_3}. The dropout is set to $p=0.1$. Since the ladder network is also regularized by unsupervised auxiliary tasks, reducing the influence of dropout led to better performance on the development set. For the noisy encoder (Fig. \ref{fig:ladder}), we add a Gaussian noise with variance $\sigma^2$ = 0.3 to the encoder. The hyper-parameter for the reconstruction loss is set to $\lambda_{l}=1$ (Eqs. \ref{eq:CostLadder} and \ref{eq:CostLadder_MTL}). A preliminary search on the development set showed no significant difference between $\lambda_{l}=1$, $\lambda_{l}=0.1$ and $\lambda_{l}=10$. We do not optimize the value of $\lambda_{l}$ to reduce the computational resources needed to train the system, acknowledging that better results may be possible by conducting an exhaustive search for this parameter over the development set. The \emph{mean squared error} (MSE) function is used to measure the reconstruction loss.

All our models are trained and evaluated using the \emph{concordance correlation coefficient} (CCC). The CCC maximizes the Pearson's correlation between the true and predicted values, while minimizing the difference between their means. Previous studies have shown the benefits of training with CCC as the objective function over the MSE \cite{Trigeorgis_2016, Ringeval_2015, Parthasarathy_2018_3}. All neural networks in this study are trained using the NADAM optimizer \cite{Dozat_2015}.

\section{Experimental Results}
\label{sec:results}

\subsection{Within Corpus Results}
\label{ssec:within_corpus}

\begin{table}[tb]		
	\caption{Within-corpus evaluation on the MSP-Podcast corpus. The results correspond to the CCC values achieved by different implementations of the ladder network architecture on the development and test sets.  ($\bullet$ indicates that one model is significantly better than the STL baseline; $\ast$ indicates that one model is significantly better than the MTL baseline).}
	\centering
	\begin{tabular*}{\columnwidth}{@{\extracolsep{\fill}}c||c|c|c}
		
		\hline
		\multirow{2}{*}{Task}& \multicolumn{3}{c}{Development}\\
		\cline{2-4}
		& Arousal & Valence & Dominance\\
		\hline
		\hline
		STL & 0.773 &	0.491	& 0.713 \\
		MTL & 0.782 & \textbf{0.509}	& 0.726  \\
		\hline
		Lad + L + STL & $0.793^{\bullet\ast}$ & 0.489 & $0.732^{\bullet}$ \\
		Lad + L + MTL & $\textbf{0.795}^{\bullet\ast}$ & 0.497 & $\textbf{0.736}^{\bullet}$ \\
		\hline
		Lad + UL + STL & $0.792^{\bullet\ast}$ & 0.489 & $0.733^{\bullet}$ \\
		Lad + UL + MTL & $0.792^{\bullet\ast}$ & 0.489 & $0.733^{\bullet}$ \\
		\hline
		\hline
		\multirow{2}{*}{}& \multicolumn{3}{c}{Test}\\
		\cline{2-4}
		& Arousal & Valence & Dominance\\
		\hline
		\hline
		STL &  0.743	& \textbf{0.312} & 0.670\\
		MTL & 0.745	& 0.293 &	0.671 \\
		\hline
		Lad + L + STL & $0.765^{\bullet\ast}$  & 0.303 & 0.678 \\
		Lad + L + MTL & $0.763^{\bullet\ast}$ & 0.293 & $0.690^{\bullet\ast}$ \\
		\hline
		Lad + UL + STL &  $\textbf{0.770}^{\bullet\ast}$ & 0.301 & $\textbf{0.700}^{\bullet\ast}$\\
		Lad + UL + MTL & $\textbf{0.770}^{\bullet\ast}$ & 0.301 & $\textbf{0.700}^{\bullet\ast}$\\
		\hline
	\end{tabular*}
	\label{tab:results}
\end{table}

\begin{figure}[tb]
	\centering
	\subfigure[Arousal]
	{
		\includegraphics[width=0.95\columnwidth]{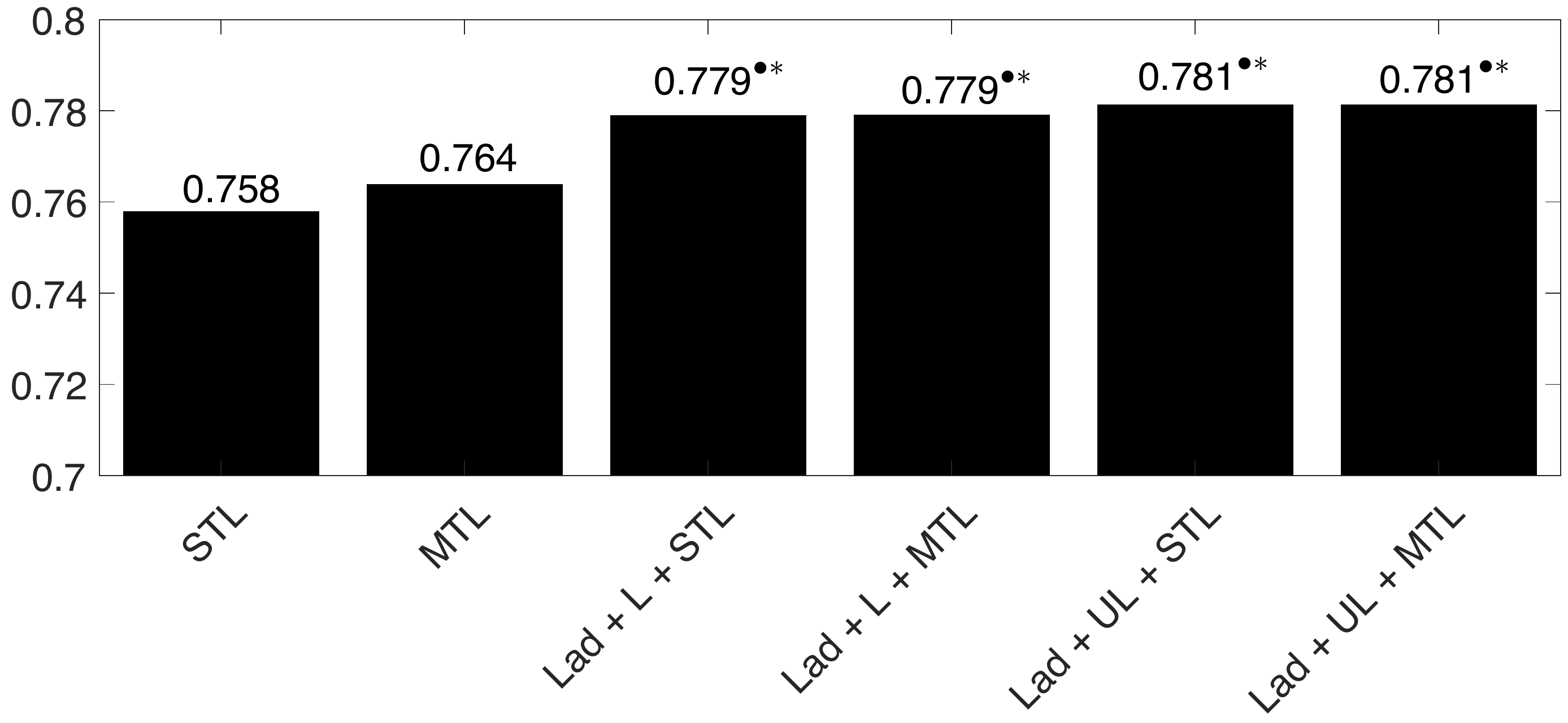}
		\label{fig:arousal_hld}
	}
	\subfigure[Valence]
	{
		\includegraphics[width=0.95\columnwidth]{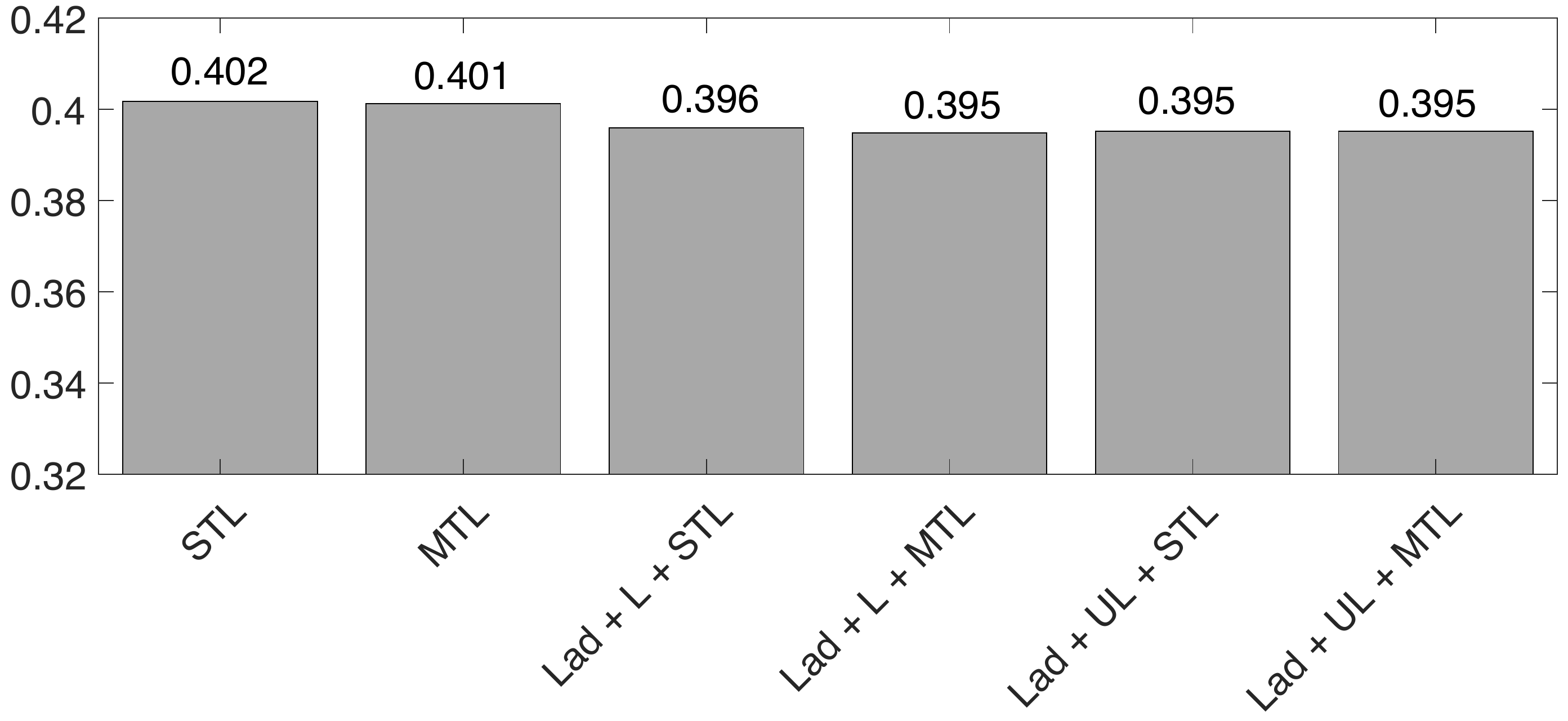}
		\label{fig:valence_hld}
	}
	\subfigure[Dominance]
	{
		\includegraphics[width=0.95\columnwidth]{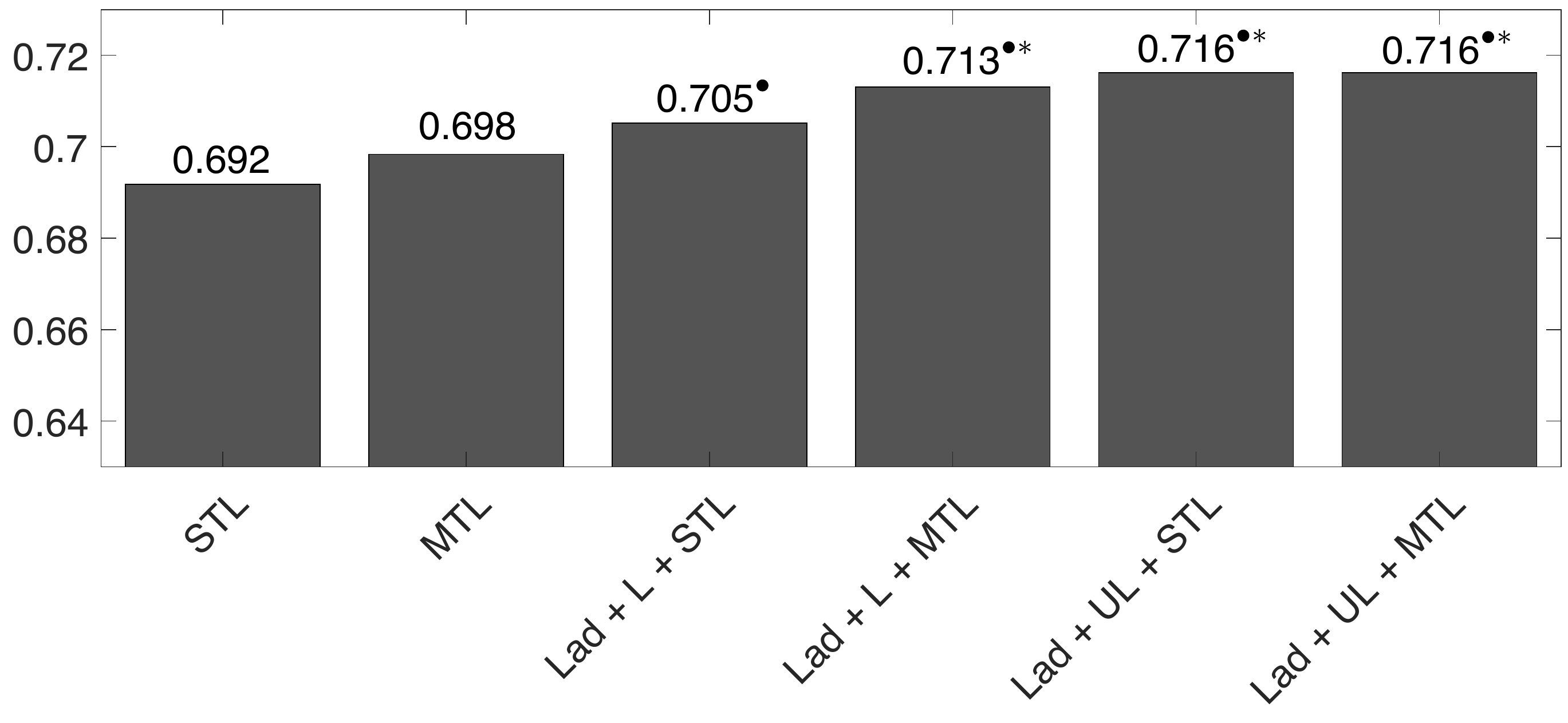}
		\label{fig:dominance_hld}
	}
	\
	\caption{Within-corpus evaluation on the MSP-Podcast corpus using sentence-level features (HLDs). The figures report the mean CCC values obtain in the development and test sets   ($\bullet$ indicates that one model is significantly better than the STL baseline; $\ast$ indicates that one model is significantly better than the MTL baseline).}
	\label{fig:results_hld}
\end{figure}

The experimental evaluation in this section analyzes the power of the proposed ladder network systems for within corpus experiments in the MSP-Podcast corpus. The systems are trained and tested on the MSP-Podcast corpus using the ComParE feature sets (Section \ref{ssec:features}). We analyze the performance in terms of CCC for arousal, valence and dominance. In this section, we report and compare the performance of our models on the development and test sets to evaluate the generalization of our approach (Table \ref{tab:results}). The development set includes the best performance, per model, across epochs obtained on this set. We compare the CCC scores of the proposed models against the baselines, asserting whether the differences in performance are statistically significant using the Fisher Z-transformation test (one-tailed z-test, $p$-value$<$0.05).

On the development set, Table \ref{tab:results} shows that the best performing systems for ladder network architectures are significantly better than the STL baseline for arousal and dominance. For these emotional attributes, the best performance is achieved by the ladder network implemented with MTL with only labeled data. 

The results on the test set are very consistent with the trends observed in the development set, demonstrating the generalization of the models (Table \ref{tab:results}). For arousal, the results of the ladder network frameworks are statistically significantly better than the results achieved by both baseline methods. For dominance, the ladder network architectures trained with labeled and unlabeled data lead to statistically significant improvements over both baseline frameworks. The frameworks trained with unlabeled data give the best performance for both arousal and dominance. Under this setting, the ladder network truly utilizes the abundant unlabeled data and generalizes to unseen data. Table \ref{tab:results} shows that for within corpus evaluations, the baseline methods achieve better results for valence. We will show in Section \ref{ssec:cross_corpora} that this is not the case for cross corpus evaluations, where our proposed ladder network architectures achieve better performance than the the baseline methods for all the emotional attributes. 

We calculate the average performance between the development and test evaluations for each of the methods, to visualize the general trends in the results. Figure \ref{fig:results_hld} shows the results, where statistically significant improvements over the baseline methods are denoted with symbols on top of the bars. Overall, we achieve relative gains of 3.0\% for arousal, and 3.5\% for dominance using the proposed architectures over the STL method.

\begin{table}[tb]		
	\caption{Cross-corpus evaluation where the models are trained on the MSP-Podcast corpus and tested  on either the USC-IEMOCAP or the MSP-IMPROV corpora. The table reports the average CCC values across folds and the standard deviation. \emph{WC Baseline} corresponds to the within-corpus baseline.  ($\bullet$ indicates that one model is significantly better than the STL baseline; $\ast$ indicates that one model is significantly better than the MTL baseline).}
	\centering
	\fontsize{7.8}{11}\selectfont
	\begin{tabular*}{0.99\columnwidth}{@{\extracolsep{\fill}}l@{\hspace{0.1cm}}|@{\hspace{0.1cm}}l@{\hspace{0.1cm}}|@{\hspace{0.1cm}}l@{\hspace{0.1cm}}|@{\hspace{0.1cm}}l}
		
		\hline
		\multirow{2}{*}{Task}& \multicolumn{3}{c}{IEMOCAP}\\
		\cline{2-4}
		& Arousal & Valence & Dominance \\
		\hline
		\hline
		STL & 0.560 $\pm$ 0.122 & 0.135 $\pm$ 0.070 & 0.378 $\pm$ 0.103 \\
		MTL & 0.584 $\pm$ 0.078 & 0.144 $\pm$ 0.067 & 0.370 $\pm$ 0.097 \\
		\hline
		Lad + L + STL & $ 0.590 \pm 0.074 ^{\bullet\ast}$ & $0.154 \pm 0.052^{\bullet}$ & $0.391 \pm 0.107 ^{\bullet\ast} $\\
		Lad + L + MTL & $0.589 \pm 0.065^{\bullet}$ & 0.141 $\pm$ 0.056 & $0.408 \pm 0.103 ^{\bullet\ast}$ \\
		\hline
		Lad + UL + STL & $0.603 \pm 0.043 ^{\bullet\ast}$ & 0.092 $\pm$ 0.071 & $0.476 \pm 0.076 ^{\bullet\ast}$ \\
		Lad + UL + MTL & $0.623 \pm 0.036 ^{\bullet\ast}$  & $0.235 \pm 0.056 ^{\bullet\ast}$ & $0.441 \pm 0.086 ^{\bullet\ast}$ \\
		\hline
		\textit{WC Baseline} & 0.661 $\pm$ 0.051 & 0.487 $\pm$ 0.044 & 0.512 $\pm$ 0.055 \\
		\hline
		\multirow{2}{*}{}& \multicolumn{3}{c}{MSP-IMPROV}\\
		\cline{2-4}
		& Arousal & Valence & Dominance \\
		\hline
		\hline
		STL & 0.471 $\pm$ 0.112 & 0.235 $\pm$ 0.078 & 0.440 $\pm$ 0.134\\
		MTL & 0.442 $\pm$ 0.116 & 0.231 $\pm$ 0.082 & 0.449 $\pm$ 0.128\\
		\hline
		Lad + L + STL & $0.490 \pm 0.108^{\ast}$ & $0.287 \pm 0.075^{\bullet\ast} $ & 0.436 $\pm$ 0.130\\
		Lad + L + MTL & $0.480 \pm 0.107^{\ast}$ & $0.293 \pm 0.073^{\bullet\ast} $ & $0.464 \pm 0.123^{\bullet\ast}$\\
		\hline
		Lad + UL + STL & $0.547 \pm 0.094 ^{\bullet\ast}$ & $0.349 \pm 0.087 ^{\bullet\ast}$  & $0.463 \pm 0.096 ^{\bullet\ast}$ \\
		Lad + UL + MTL & $0.547 \pm 0.094 ^{\bullet\ast}$ & $0.328 \pm 0.091 ^{\bullet\ast}$ & $0.463 \pm 0.096 ^{\bullet\ast}$ \\
		\hline
		\textit{WC Baseline} &0.599 $\pm$ 0.112 & 0.408 $\pm$ 0.090 & 0.471 $\pm$ 0.098 \\
		\hline
	\end{tabular*}
	\label{tab:cross_results}
\end{table}

\begin{figure}[t]
	\centering
	\subfigure[Arousal]
	{
		\includegraphics[width=\columnwidth]{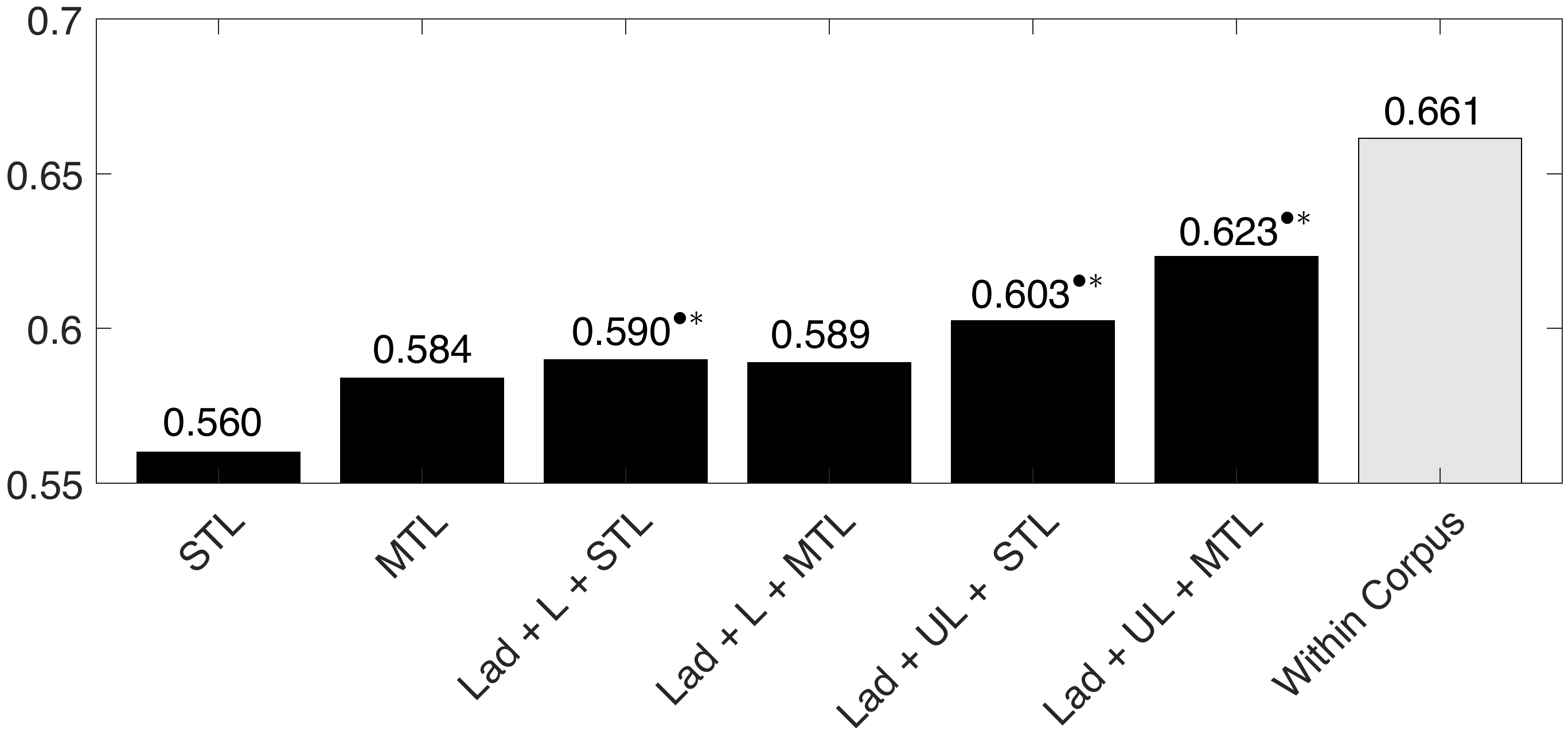}
		\label{fig:arousal_iemocap}
	}
	\subfigure[Valence]
	{
		\includegraphics[width=\columnwidth]{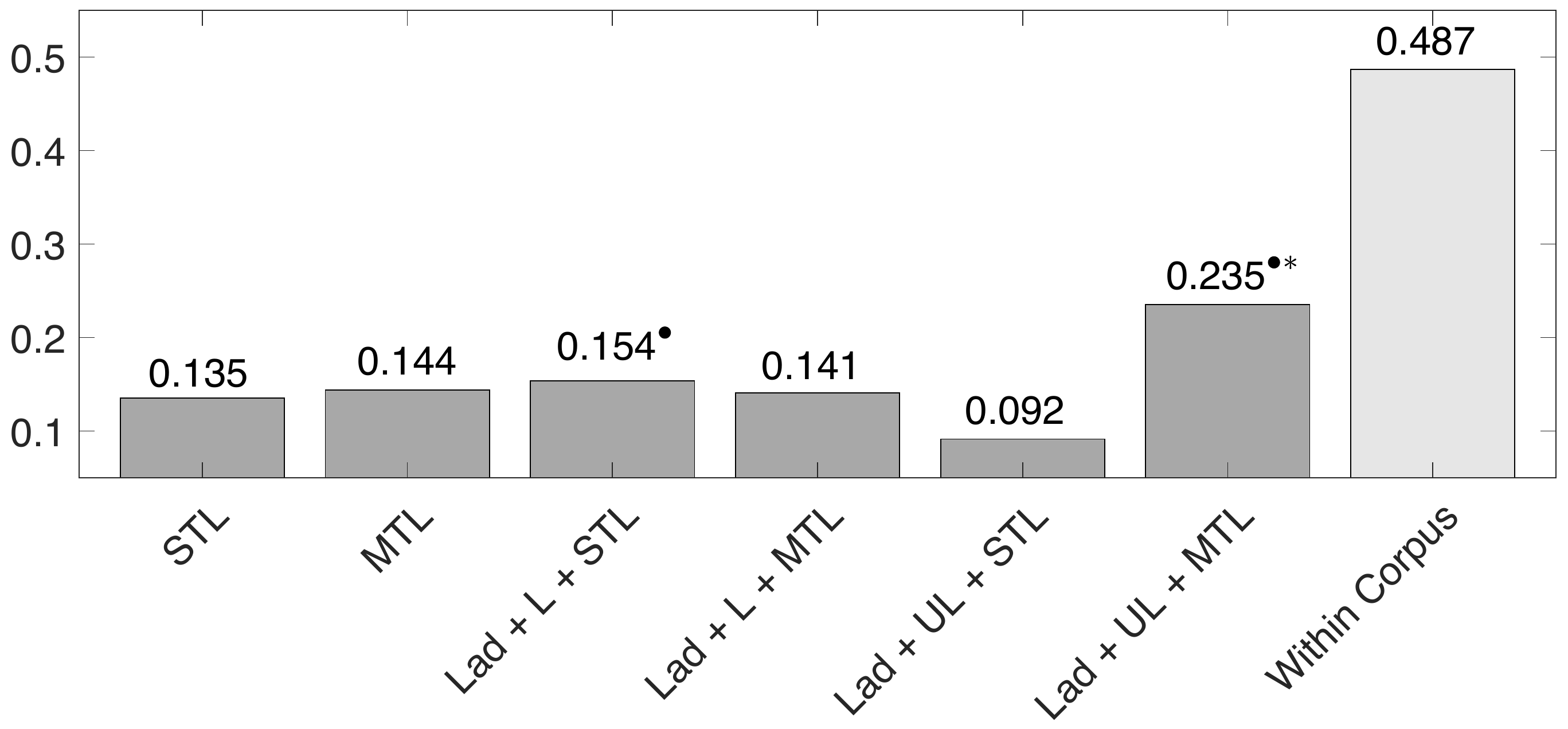}
		\label{fig:valence_iemocap}
	}
	\subfigure[Dominance]
	{
		\includegraphics[width=\columnwidth]{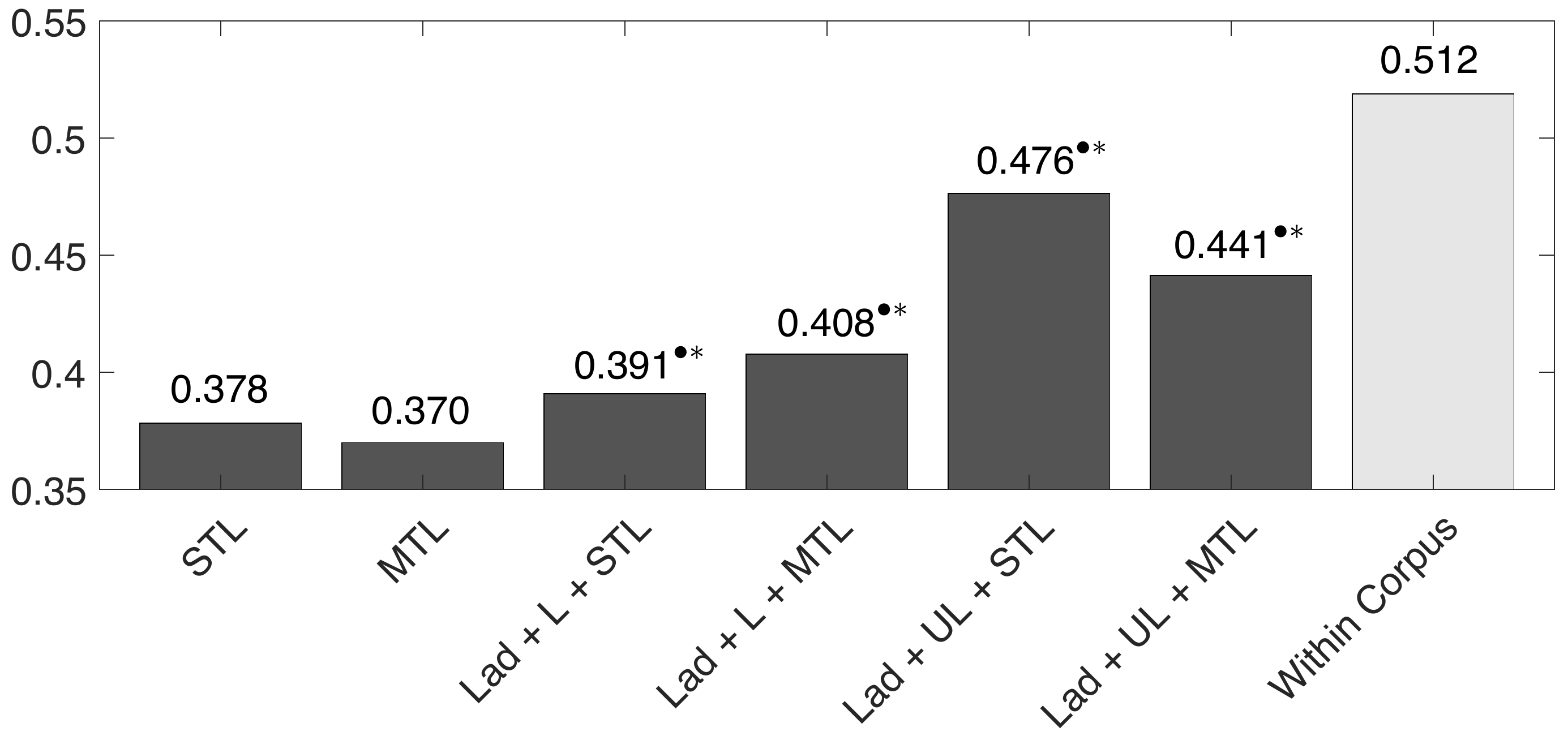}
		\label{fig:dominance_iemocap}
	}
	\
	\caption{Cross-corpus evaluation, when the models are tested on the USC-IEMOCAP corpus. The figure reports the average CCC values across folds ($\bullet$ indicates that one model is significantly better than the STL baseline; $\ast$ indicates that one model is significantly better than the MTL baseline).}
	\label{fig:results_iemocap}
\end{figure}

\begin{figure}[t]
	\centering
	\subfigure[Arousal]
	{
		\includegraphics[width=\columnwidth]{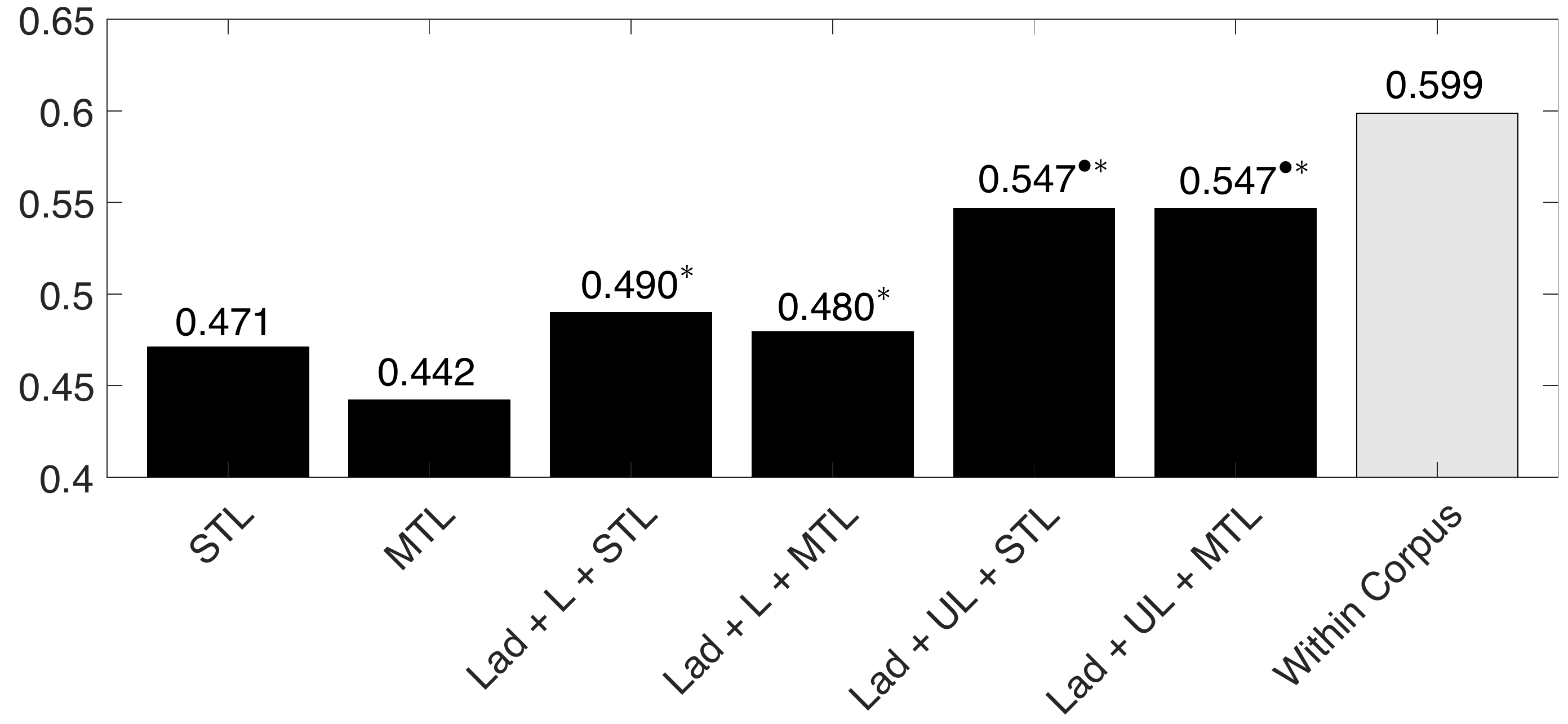}
		\label{fig:arousal_improv}
	}
	\subfigure[Valence]
	{
		\includegraphics[width=\columnwidth]{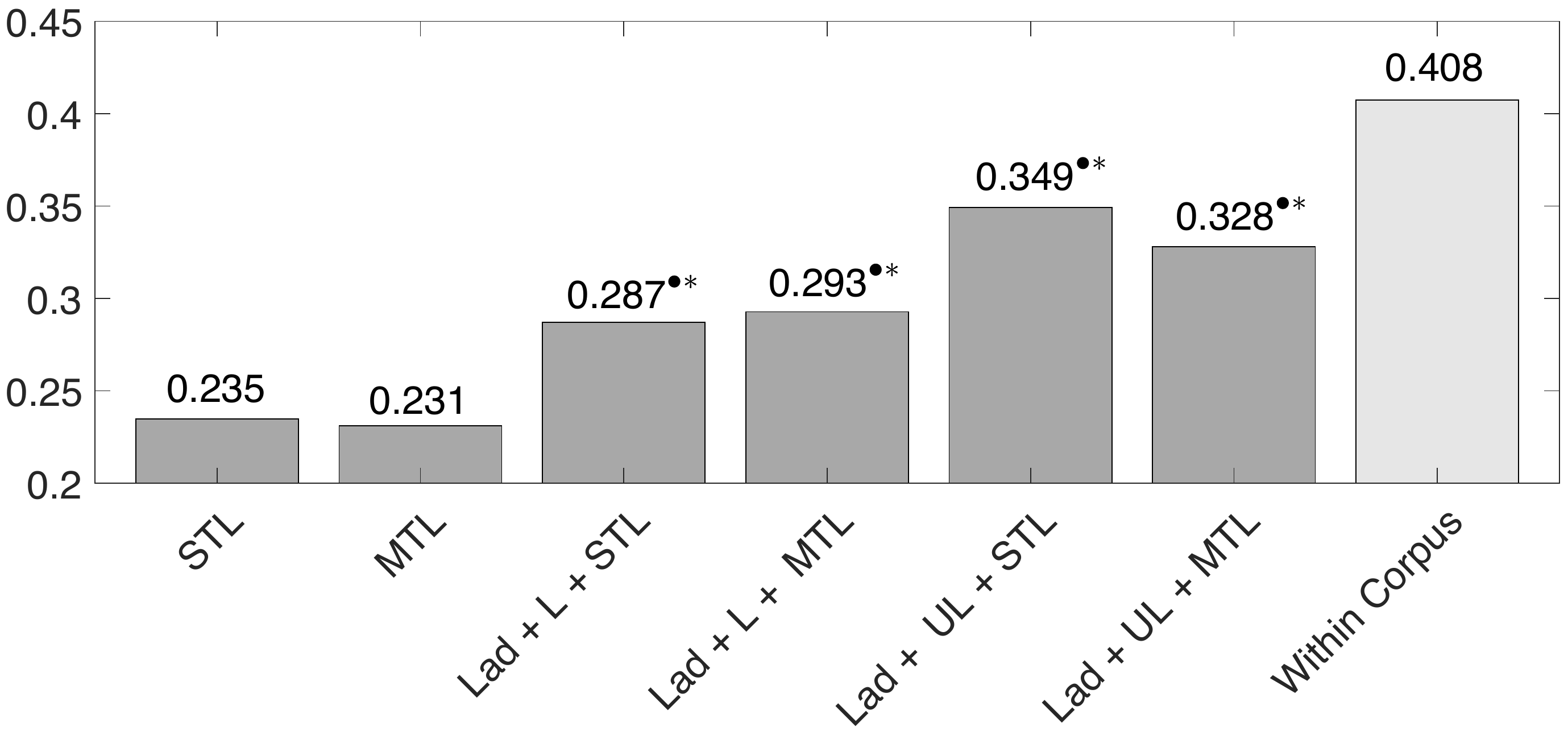}
		\label{fig:valence_improv}
	}
	\subfigure[Dominance]
	{
		\includegraphics[width=\columnwidth]{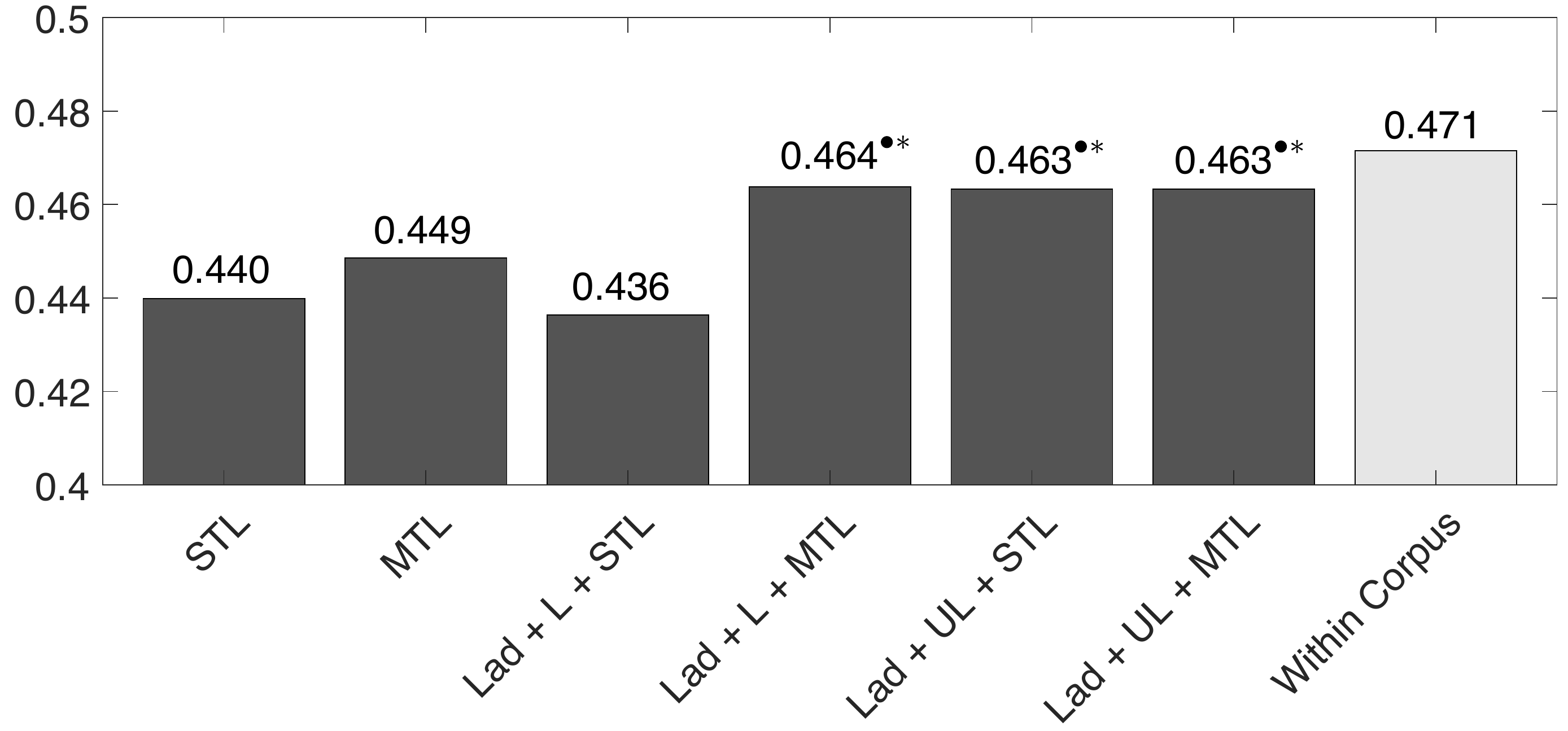}
		\label{fig:dominance_improv}
	}
	\
	\caption{Cross-corpus evaluation, when the models are tested on the MSP-IMPROV corpus. The figure reports the average CCC values across folds ($\bullet$ indicates that one model is significantly better than the STL baseline; $\ast$ indicates that one model is significantly better than the MTL baseline).}
	\label{fig:results_improv}
\end{figure}

\subsection{Cross Corpus Results}
\label{ssec:cross_corpora}

\begin{figure*}[tb]
	\centering
	\includegraphics[width=2\columnwidth]{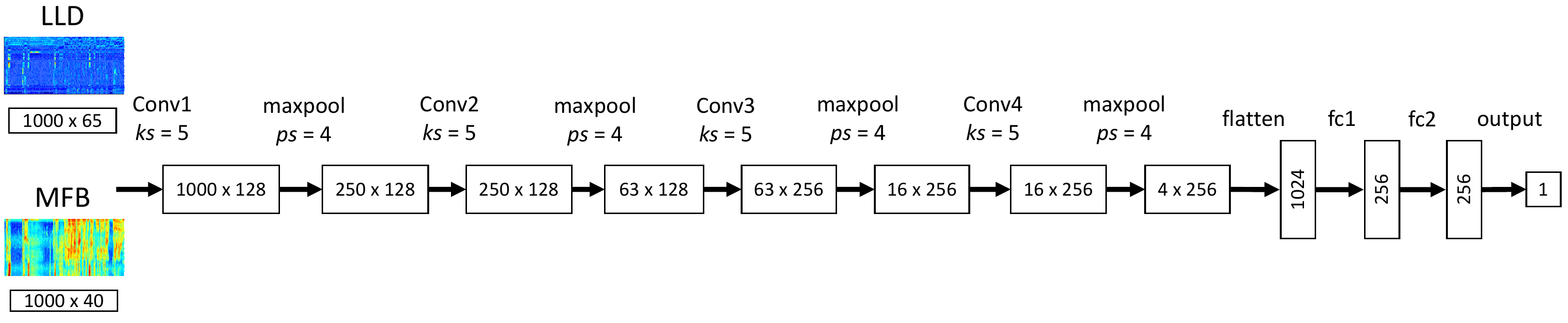}
	\caption{CNN architecture to predict emotional attributes with frame-level features (LLDs or MFB). The architecture contains 4 blocks of 1D convolutional layer followed by a 1D maxpooling layer. After flattering the last convolution layer, the network includes two fully connected layers and the output prediction layer (\emph{ks}: kernel size, \emph{ps}: pool size and \emph{fc}: fully connected).}
	
	\label{fig:cnn_structure}
\end{figure*}

This study also explores the generalization of the proposed ladder network with cross-corpus experiments.  Specifically, we train the models on the MSP-Podcast corpus maximizing performance on its development set. The models are then tested on either the USC-IEMOCAP corpus or MSP-IMPROV corpus. We compare the results with within corpus evaluations using the STL framework, where the models are trained and tested with data from the same corpus (\emph{Within-corpus (WC) Baseline} in Table \ref{tab:cross_results}). For the within corpus evaluation, the USC-IEMOCAP and MSP-IMPROV corpora are divided into speaker independent partitions. The results are reported across all the test partitions. For consistency, the results for the ladder networks are also estimated for each partition, reporting the average across folds. 

We train the ladder network architectures introduced in Section \ref{ssec:description} using labeled and unlabeled data. For the labeled setting, we use samples only from the MSP-Podcast corpus. For the unlabeled setting (\emph{UL} in Table \ref{tab:cross_results}), we assume we have access to the samples from the target corpus. We include the target corpus for the unsupervised reconstruction using the autoencoder. Notice that we use the speech recordings, without the emotional labels. Since the  emotional attributes in the MSP-Podcast and the target corpora are annotated on different scales, we transform the attribute scores of the MSP-Podcast corpus to match the scales of the target corpora using an affine transformation. We report the mean and standard deviation over all the test partitions. We compare the CCC values obtained by the ladder networks with the results from the baselines, testing their significance with the one-tailed, matched-paired t-test asserting significance at $p$-value$<$0.05. Table \ref{tab:cross_results} describes the results for the cross-corpus experiments.  Figures \ref{fig:results_iemocap} (USC-IEMOCAP) and \ref{fig:results_improv} (MSP-IMPROV) illustrate the mean performance across test partitions.

First, we discuss the results for the USC-IEMOCAP database. Under the fully labeled setting, the ladder network systems achieve significant improvements over the STL baseline for arousal and dominance. Additionally, the systems significantly improve the performance for dominance compared to the MTL baseline. For valence, we achieve significant gains over the STL baseline with the \emph{Ladder + L + STL} model. With unlabeled data from the USC-IEMOCAP corpus (UL setting), we obtain significant gain over the baselines. The systems perform significantly better than the baselines for all three emotional attributes. We observe relative gains up to 11.3\% for arousal, 74.1\% for valence, and 25.9\% for dominance over the STL baseline (Figure \ref{fig:results_iemocap}). The CCC values for these systems are closer to the results obtained by the within-corpus baseline. The significant gains reported in this section show the potential of the ladder network architecture, especially when unlabeled data from the target corpus is available.

We observe similar results in the evaluation on the MSP-IMPROV database, where most of the architectures using ladder network achieve significant improvements in the CCC values over the STL and MTL baselines. For valence, the proposed architectures perform significantly better than both baselines. For arousal and dominance, the use of unlabeled data leads to statistically significant improvements over the STL and MTL baselines. Figure \ref{fig:results_improv} shows that the inclusion of unlabeled data from the target corpus greatly improves the performance of the ladder network architectures, achieving CCC scores that are closer to the within-corpus baseline. Under this setting, the proposed systems are significantly better than the baselines for all three emotional attributes. Overall, the \emph{Lad + UL +MTL} architecture achieves relative gains of 16.1\% for arousal, 40\% for valence, and 5.5\% for dominance over the STL baseline. These results demonstrate the real benefits of the ladder network architecture, which generalizes better in cross corpus SER problems.

\subsection{Results with Frame-Level Features}
\label{ssec:cnn}

Conventionally, SER problems are formulated using sentence-level features over short speech segments. Previous studies often rely on statistics estimated over LLDs, where popular examples include the feature sets proposed for the paralinguistic challenges at Interspeech \cite{Schuller_2009, Schuller_2013}. An alternative approach is to directly use a sequence of features extracted at the frame-level over short segments (e.g., 40 ms). We refer to these features as frame-level features. With the advancements of DNNs, many recent techniques for SER have focused on frame-level dynamic features. A popular SER option among these frameworks is the use of CNNs to learn high-level feature representations from the frame-level representations. Cummins \etal \cite{Cummins_2017} borrowed successful CNN architectures from the computer vision domain by treating speech spectrograms as images. Mao \etal \cite{Mao_2014} performed SER in a two step approach using a CNN architecture on frame-level features. The first step learned features from unlabeled data and a sparse autoencoder. These features were then used for the recognition task. Trigeorgis \etal \cite{Trigeorgis_2016} proposed a CNN architecture to perform end-to-end SER that took raw speech waveforms as inputs. They showed high correlation between the learned features and hand-crafted features which have been used in previous works (e.g loudness, mean fundamental frequency). Neumann and Vu \cite{Neumann_2017} proposed an attention based convolutional neural network for emotion recognition. An attention layer was used at the final feature representation from the CNN that performed a weighted pooling of the features from different time frames. Yang and Hirschberg \cite{Yang_2018} predicted arousal and valence using CNNs trained on spectrogram inputs. Finally, Aldeneh and Provost \cite{Aldeneh_2017} proposed to train 1-D CNNs on mel-filter bank energies to capture regional saliency for emotion recognition. Following these previous works, our study also examines the effect of our system using CNNs on frame-level features, demonstrating that the proposed ladder network architecture can also be implemented with these features. 

Most previous frameworks use either frame-level features (e.g., MFB) or audio waveforms to learn discriminative features for the task at hand. Such methods enable end-to-end learning, where the features and the classification or regression tasks are jointly learned during training. Following this formulation, this study explores the use of the proposed ladder networks with frame-level features. We consider two alternative frame-level features: (1) using the LLDs of the ComParE feature set (65D vector -- see Sec. \ref{ssec:features}), and (2) MFB energies. Similar to previous studies, we use $n$=40 bands for the MFB \cite{Aldeneh_2017}. These models are compared with systems trained with HLDs.

Figure \ref{fig:cnn_structure} shows the proposed CNN-based architecture for frame-based features. The input to the CNN is a 65D $\times T$ matrix (ComParE LLD) or a 40D $\times T$(MFB) matrix, where $T$ is the time dimension. The CNN architecture consists of four convolutional layers followed by two \emph{fully connected} (FC) layers and a linear output layer. The convolutional layers perform 1D convolutions along the time axis with the frame-level features as the inputs for the first convolutional layer. We use a 1D max pooling layer after every convolutional layer to sequentially reduce the dimension of the time axis. We flatten the outputs from the final convolutional layer before passing them to the FC layers. While the downstream convolutional layers can deal with variable length sequences, the upstream FC layers require a fixed length input. Therefore, we fix $T$ at 1000, which corresponds to 10 seconds of speech (100 fps). We use this value, since most speech segments in the different datasets used in this study are less than 10 seconds. Segments with duration greater than 10 seconds are truncated. Sentences with duration less than 10 seconds are padded with zeros.

The analysis in this section includes only within corpus experiments on the MSP-Podcast corpus. All the parameters for the CNN architecture are optimized on the development set of the MSP-Podcast corpus. Training ladder networks with frame-level features is computationally expensive. To ease this process, we impose two constraints on the ladder networks trained with frame-level features. First, the reconstruction costs are implemented only on the two fully connected layers after the flattening layer (i.e., layers \emph{fc1} and \emph{fc2} in Fig.\ref{fig:cnn_structure}). This network is similar to the $\tau$ network suggested by Valpola \etal \cite{Valpola_2015}. Second, we do not use the entire unlabeled portion of the corpus in every epoch. Instead, we use the same number of unlabeled and labeled samples for every epoch, randomly selecting 29,440 unlabeled samples in every epoch. The STL and MTL baselines are also implemented with CNNs.

\begin{table}[tb]		
	\caption{Evaluation of ladder network with frame-level features. The results correspond to within-corpus evaluations using the MSP-Podcast corpus. The table reports CCC for different architectures using CNNs trained with either LLDs or MFB ($\bullet$ indicates that one model is significantly better than the STL baseline; $\ast$ indicates that one model is significantly better than the MTL baseline).}
	\centering
	\begin{tabular*}{\columnwidth}{@{\extracolsep{\fill}}c||c|c|c}
		
		\hline
		\multirow{2}{*}{Task}& \multicolumn{3}{c}{LLD-CNN}\\
		\cline{2-4}
		& Arousal & Valence & Dominance \\
		\hline
		\hline
		STL &  0.756 & 0.244 & 0.682 \\
		MTL &  0.759 & 0.223 & 0.684 \\
		\hline
		Lad+STL+L & $0.768^{\bullet}$ & $0.274^{\bullet \ast}$ & \textbf{0.687} \\
		Lad+MTL+L & $0.769^{\bullet}$ & $0.274^{\bullet \ast}$ & 0.681 \\
		\hline
		Lad+STL+UL &  $0.769^{\bullet}$ & $\textbf{0.279}^{\bullet \ast}$ & \textbf{0.687} \\
		Lad+MTL+UL & $\textbf{0.771}^{\bullet \ast}$ & $0.269^{\ast}$ & 0.685 \\
		\hline
		\multirow{2}{*}{}& \multicolumn{3}{c}{MFB-CNN}\\
		\cline{2-4}
		& Arousal & Valence & Dominance\\
		\hline
		\hline
		STL & 0.733 & 0.204 & \textbf{0.659} \\
		MTL & 0.738 & $\textbf{0.254}^{\bullet}$ & \textbf{0.659} \\
		\hline
		Lad+STL+L & \textbf{0.744} & 0.200 & 0.659 \\
		Lad+MTL+L & 0.741 & 0.200 & 0.659 \\
		\hline
		Lad+STL+UL & 0.743 & $0.232^{\bullet}$ & 0.655 \\
		Lad+MTL+UL & 0.740 & 0.184  & 0.656 \\

		\hline
	\end{tabular*}
	\label{tab:results_frame}
\end{table}

Table \ref{tab:results_frame} shows the results for the different systems using the CNN-based architecture trained with either LLDs or MFB features. Similar to Section \ref{ssec:within_corpus}, we evaluate the differences in CCC values using the Fisher Z-transformation (one-tailed z-test, $p$-value$<$0.05). When the CNNs are trained with LLDs, we observe that the ladder networks provide significant gains over the baseline for arousal (STL) and valence (STL, MTL). For valence, the proposed architectures provide relative gains up to 14.3\% on the test set. For dominance, the models achieve similar performance to the baselines, where the differences are not statistically significant. When the CNNs are trained with MFB, we observe similar performance. We observe statistically significant improvements over the STL baseline only for valence using the \emph{Lad+STL+UL} network. We expect that a better result can be achieved if the reconstruction loss is implemented to also include the convolutional layers.

\begin{figure}[tb]
	\centering
	\subfigure[Arousal]
	{
		\includegraphics[width=0.8\columnwidth]{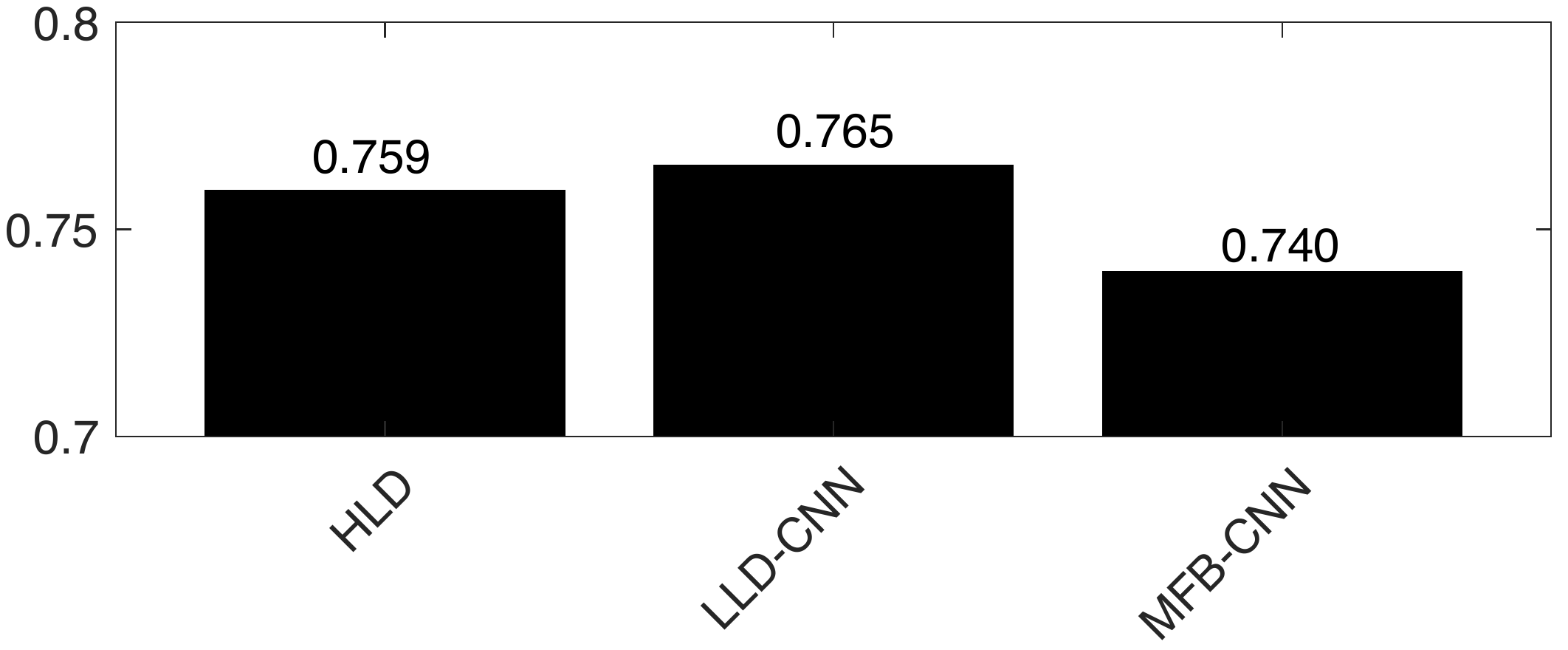}
		\label{fig:arousal}
	}
	\subfigure[Valence]
	{
		\includegraphics[width=0.8\columnwidth]{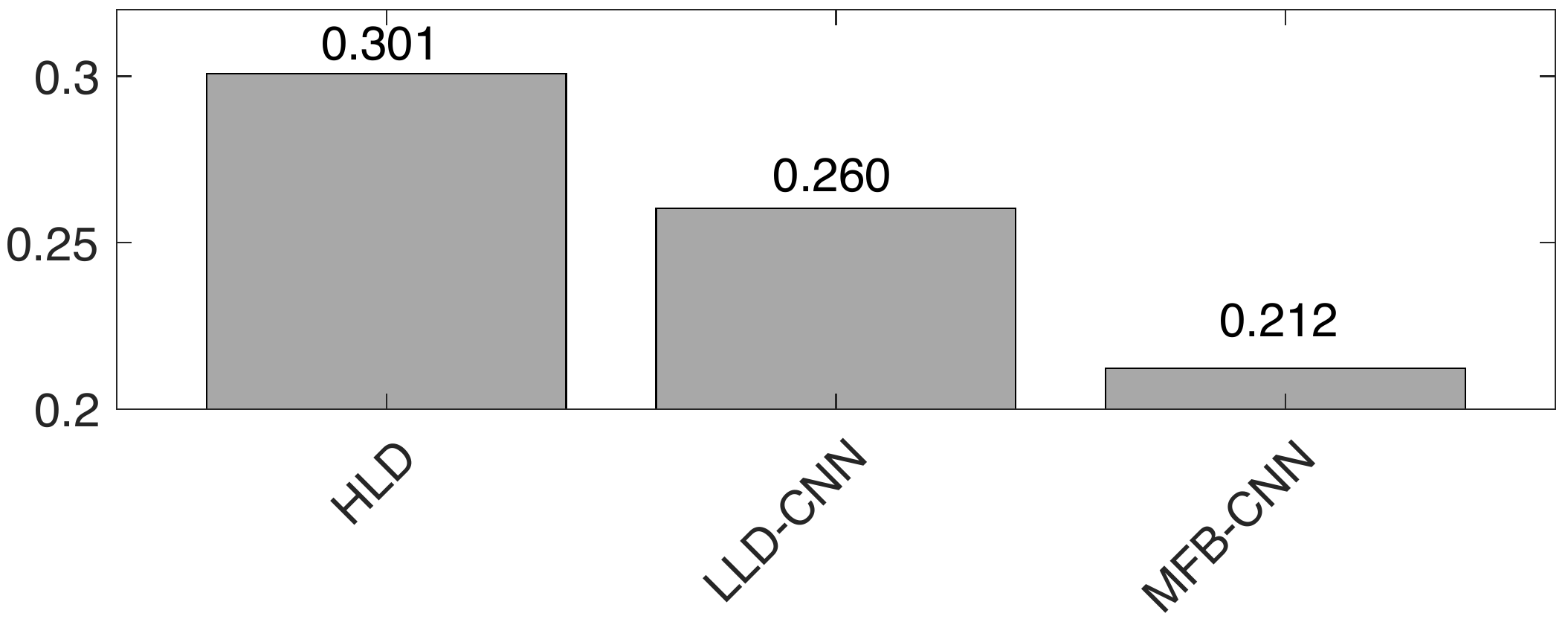}
		\label{fig:valence}
	}
	\subfigure[Dominance]
	{
		\includegraphics[width=0.8\columnwidth]{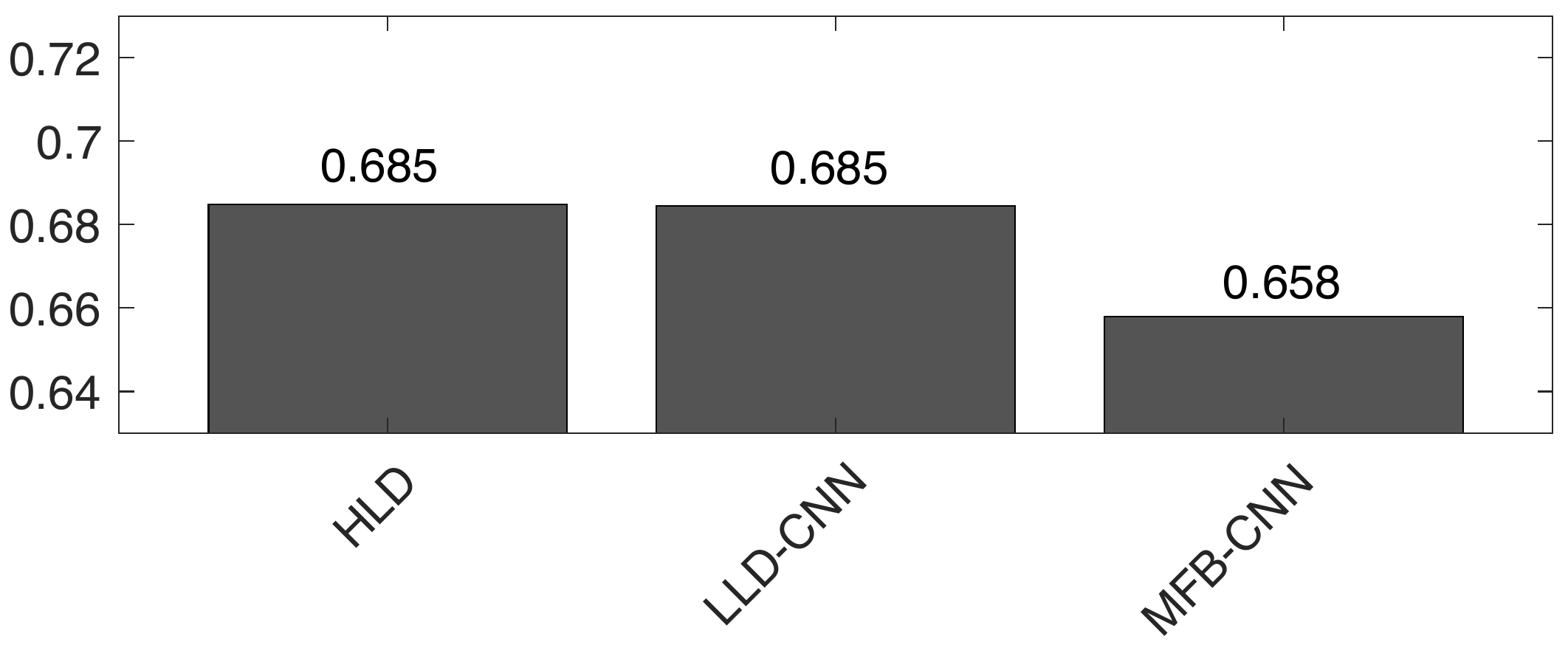}
		\label{fig:dominance}
	}
	\
	\caption{Evaluation of ladder network with frame-level features. Mean CCC values across different architectures for sentence-level features (HLD) and frame-level features (LLDs and MFB). Frame-level features are implemented with CNN using the architecture described in Figure \ref{fig:cnn_structure}.}
	\label{fig:results_comparison}
\end{figure}

Finally, we also compare the overall results of the models trained with sentence-level features (HLD) and frame-level features (CNN-LLD, CNN-MFB). Figure \ref{fig:results_comparison} shows the average CCC results across architectures, including the baselines. For arousal and dominance, we observe similar performance for systems trained with either the HLDs (sentence-level features), or the CNN-LLD (frame-label features). In contrast, the system trained with sentence-level features achieves better results for valence. The results are consistently lower when using MFB. MFB features only provide spectral information, while the LLDs and HLDs also provide prosodic and voice quality information, which are important cues for SER problems.

\section{Conclusions}
\label{sec:conclusion}

This study proposed the use of ladder network in speech emotion recognition. The approach combines the unsupervised auxiliary task of reconstructing intermediate feature representations, with the primary task of predicting emotional attributes. The unsupervised nature of the auxiliary task eases the pressure on the expensive emotion labeling process by leveraging unlabeled data from the source domain. The unsupervised auxiliary task reconstructs the input and the intermediate feature representations through a denoising autoencoder. The ladder networks contain skip connections between the noisy encoder and the decoder, allowing the higher layers of the encoder to learn discriminative representations. Different implementations of the proposed system were evaluated in within corpus evaluations and cross-corpus evaluations. In the within-corpus evaluations, we analyzed the benefits of the proposed architectures over competitive STL and MTL baselines, showing significant improvements for arousal and dominance. In the cross-corpus evaluations, the models were trained on the MSP-Podcast corpus and evaluated on the USC-IEMOCAP and MSP-IMPROV corpora. The results indicated significant gains when using the proposed models, underlying the generalization power of the ladder networks. The improvements were particularly high when using unlabeled data from the target domain, exploiting all the benefits of the proposed architecture. Finally, the study analyzed the performance of the proposed architecture for different feature inputs. We showed that we can achieve similar performance with a CNN-based implementation trained on frame-level features.

Based on the cross-corpus results, our future work will explore the use of the representations learned by the ladder networks as inputs for emotion recognition tasks in general. We will also explore the ladder network architecture for emotion recognition from other modalities such as video and image. Previous studies have shown the difficulty of predicting valence from acoustic cues \cite{Sridhar_2018, Busso_2012}. Our recent study have shown that acoustic cues for valence are highly speaker dependent, where the networks require higher regularization  \cite{Sridhar_2018}. Our future research direction will use these findings to improve the ladder network architectures for predicting valence scores. Finally, we aim to improve the performance of the ladder network architecture with frame-level features, paying special attention to valence, extending the scope of our data-driven speech emotion recognition systems.
\section*{Acknowledgment}
This study was funded by the National Science Foundation (NSF) CAREER grant IIS-1453781.

\ifCLASSOPTIONcaptionsoff
  \newpage
\fi



%
\bibliographystyle{IEEEtran}
\bibliography{reference}

%

\vspace{-1.2cm}




	\vspace{-0.3cm}
	\begin{IEEEbiography}[{\includegraphics[width=1in,height=1.25in,clip,keepaspectratio]{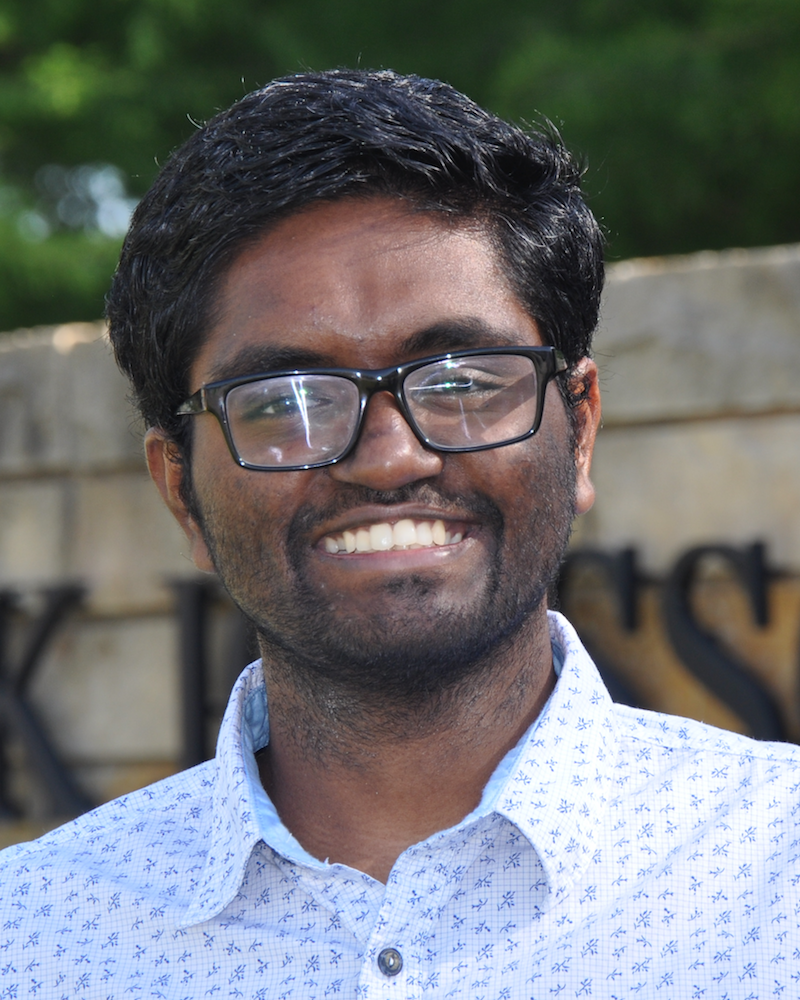}}]{Srinivas Parthasarathy} received his BS degree in degree in Electronics and Communication Engineering from College of Engineering Guindy, Anna University, Chennai, India (2012) and MS (2014) and PhD (2019) degrees in Electrical Engineering from the University of Texas at Dallas - UT Dallas. During the academic year 2011-2012, he attended as an exchange student The Royal Institute of Technology (KTH), Sweden. He is an applied scientist at Amazon Research. At UT Dallas, he was a member of the Multimodal Signal Processing (MSP) laboratory. He received the Ericsson Graduate Fellowship during 2013-2014. He has been a research intern at Amazon, Microsoft Research and Bosch Research and Training Center. His research interest includes affective computing, human machine interaction, machine learning and digital signal processing.
	\end{IEEEbiography}

\vspace{-1.0cm}
\begin{IEEEbiography}[{\includegraphics[width=1in,height=1.25in,clip,keepaspectratio]{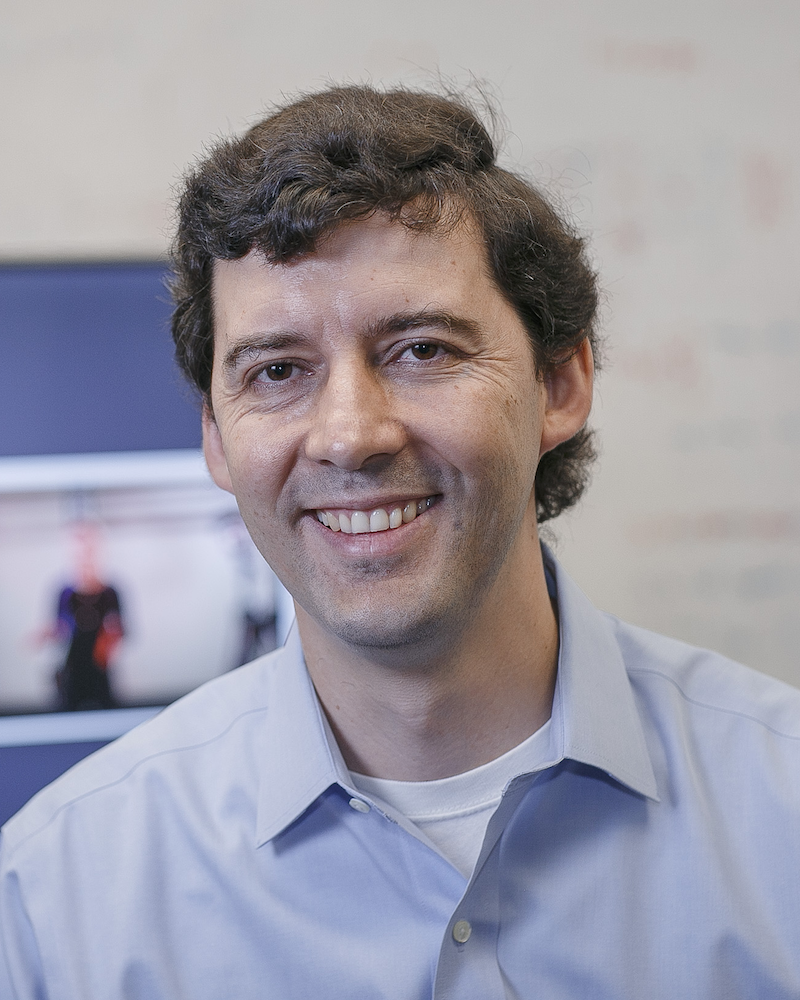}}]{Carlos Busso} 
(S'02-M'09-SM'13) received the BS and MS degrees with high honors in electrical engineering from the University of Chile, Santiago, Chile, in 2000 and 2003, respectively, and the PhD degree (2008) in electrical engineering from the University of Southern California (USC), Los Angeles, in 2008. He is an associate professor at the Electrical Engineering Department of The University of Texas at Dallas (UTD). He was selected by the School of Engineering of Chile as the best electrical engineer graduated in 2003 across Chilean universities. At USC, he received a provost doctoral fellowship from 2003 to 2005 and a fellowship in Digital Scholarship from 2007 to 2008. At UTD, he leads the Multimodal Signal Processing (MSP) laboratory [http://msp.utdallas.edu]. He is a recipient of an NSF CAREER Award. In 2014, he received the ICMI Ten-Year Technical Impact Award. In 2015, his student received the third prize IEEE ITSS Best Dissertation Award (N. Li). He also received the Hewlett Packard Best Paper Award at the IEEE ICME 2011 (with J. Jain), and the Best Paper Award at the AAAC ACII 2017 (with Yannakakis and Cowie). He is the co-author of the winner paper of the Classifier Sub-Challenge event at the Interspeech 2009 emotion challenge. His research interest is in human-centered multimodal machine intelligence and applications. His current research includes the broad areas of affective computing, multimodal human-machine interfaces, nonverbal behaviors for conversational agents, in-vehicle active safety system, and machine learning methods for multimodal processing. His work has direct implication in many practical domains, including national security, health care, entertainment, transportation systems, and education. He was the general chair of ACII 2017. He is a member of ISCA, AAAC, and ACM, and a senior member of the IEEE.
\end{IEEEbiography}

\end{document}